\documentclass[12pt]{iopart}
\usepackage{iopams}  
\usepackage{amsbsy}

\usepackage{cite}

\usepackage{graphicx}
\usepackage{dcolumn}
\usepackage{bm}

\usepackage[colorlinks=true,linkcolor=blue,urlcolor=blue,citecolor=blue]{hyperref}

\usepackage[normalem]{ulem}
\newcommand{\stkout}[1]{\ifmmode\text{\sout{\ensuremath{#1}}}\else\sout{#1}\fi}

\newcommand{\bsla}{\boldsymbol{\lambda}}
\newcommand{\kT}{k_{\rm B}T}
\newcommand{\md}{{\rm d}}
\newcommand{\mcD}{\mathcal{D}}
\newcommand{\ti}{{\Delta t_i}}           
\newcommand{\tz}{{\Delta t_0}}           
\newcommand{\tgen}{{\Delta t}}
\newcommand{\tND}{\widetilde{\tgen}}

\begin{document}

\title[Optimal discrete control]{Optimal discrete control: minimizing dissipation in discretely driven nonequilibrium systems}

\author{Steven J.~Large$^*$ and David A.~Sivak$^{\dagger}$}

\address{Department of Physics, Simon Fraser University, Burnaby, BC, Canada}
\eads{\mailto{$^*$slarge@sfu.ca}, \mailto{$^{\dagger}$dsivak@sfu.ca}}
\vspace{10pt}
\begin{indented}
\item[]\today
\end{indented}

\begin{abstract}
Microscopic machines utilize free energy to create and maintain out-of-equilibrium organization in virtually all living things. Often this takes the form of converting the free energy stored in nonequilibrium chemical potential differences into useful work, via a series of reactions involving the binding, chemical catalysis, and unbinding of small molecules. Such chemical reactions occur on timescales much faster than the protein conformational rearrangements they induce. Here, we derive the energetic cost for driving a system out of equilibrium via a series of such effectively instantaneous (and hence discrete) perturbations. This analysis significantly generalizes previously established results, and provides insight into qualitative, as well as quantitative, aspects of finite-time, minimum-dissipation discrete control protocols. We compare our theoretical formalism to an exactly solvable model system and also demonstrate the dissipation reduction achievable in a simple multistable model for a discretely driven molecular machine.
\end{abstract}

%

\section{\label{sec:introduction}Introduction}
At all scales, biological systems exhibit a striking degree of organization and coordination. The continuous flow of information, energy, and material within and between biological cells preserves the ordered structure necessary for their proper functioning. At microscopic scales, a variety of molecular machines are largely responsible for maintaining this order,	performing a broad range of intracellular tasks~\cite{howard}. For instance, the rotary ${\rm F}_{\rm o}{\rm F}_1$ ATP-synthase motor produces the cellular energy currency ATP~\cite{yoshida_2001}, whereas transport motors such as kinesin, myosin, and dynein are responsible for the directed trafficking of material within the cell~\cite{hirokawa_2009,sellers_2000}. 
	
Operationally, molecular machines function by coupling the free energy stored in nonequilibrium environmental conditions (often imbalances of chemical potential) to mechanical motion. The ${\rm F}_1$ ATPase motor, for instance, can operate at speeds up to $\sim$350 rotations per second~\cite{ueno_2004}, presumably far from thermodynamic equilibrium, by coupling rotational torque of a central crankshaft to the hydrolysis or synthesis of ATP. 
	
A variety of single-molecule experimental techniques have made possible the detailed study of these molecular machines~\cite{liu_2014,yoshida_2001,ueno_2004,yasuda_2001,berndsen_2014}, but their out-of-equilibrium operation complicates a theoretical understanding of energetic flows into, within, and out of these systems. Under the hypothesis that selective pressures favor efficient cellular machinery~\cite{bialekBook,niven_2008}, a theory which elucidates how energetic flows depend on operational parameters promises to deepen our understanding of the fundamental operational constraints facing evolved molecular machines. In practice, this would aid in the creation of \emph{de novo} molecular machines, perhaps accelerating their implementation for next-generation nanomedicine~\cite{chen_2015,peng_2017}
	
Much previous work on the properties of minimum-dissipation control (applied to a variety of model systems including the erasure of a classical bit~\cite{zulkowski_2014} and the reversal of magnetization in an Ising magnet~\cite{rotskoff_2015}, among others~\cite{schmiedl_2007,GomezMarin:2008:JChemPhys,esposito_2010,zulkowski_2012,zulkowski_2015,sivak_2016}) has assumed that the system of interest is subjected to a controlling apparatus that can be manipulated in a continuous manner~\cite{sivak_2012}. While this applies well to the single-molecule experimental paradigm, many interesting microscopic systems, such as biomolecular machines, often drive their mechanical motion via a sequence of chemical reactions. The time scales of chemical reaction and mechanical response can differ by several orders of magnitude; as such, the driving process is well approximated by a series of stochastically timed discrete perturbations to a thermodynamic system, as opposed to a continuous driving process. 

Linear-response theory has been a vital tool in the investigation of nonequilibrium relaxation, from relaxation kinetics~\cite{bernasconi} to chemical dynamics~\cite{Kutz:1974em,Skinner:1978ef,Chandler:1978hh} to nonequilibrium dissipation~\cite{kubo_1966,onsager_1931_1,onsager_1931_2}. 
As a first-order theory, we sacrifice applicability far from the considered limits in order to gain tractability and generality.  Moreover, linear-response theory has proven surprisingly effective in describing dissipation, even in contexts that clearly extend beyond its regime of rigorous validity~\cite{Tafoya:2019hr}.

In this article, as a first step toward adapting the control-protocol framework to chemically driven systems, and to isolate the effect of discrete control parameter changes, we develop a theoretical framework for nonequilibrium control using deterministic discrete control parameter changes. Our central result~\eref{cost_function} quantifies the nonequilibrium energetic costs associated with discretely driving a microscopic system. In particular, we assume the system is within the linear-response regime and is subject to sufficiently weak perturbations that a low-order approximation of the energetic costs of discrete steps is warranted. Within these limitations, this framework allows for straightforward optimization of discrete driving protocols that accounts for both the effects of the size of discrete steps as well as the local relaxation times, which leads to novel characteristics of discrete protocols not observed in continuously driven systems (Fig.~\ref{fig:continuous-limit}). This work complements and generalizes previous results on the entropy production associated with discrete processes~\cite{nulton_1985}. In the continuous-driving limit, our formalism reduces to previously known results, namely the thermodynamic-length formalism introduced by Crooks~\cite{crooks_2007} and the entropy-differential metric of Burbea and Rao~\cite{burbea_1982}, and is related to the generalized friction coefficient of Sivak and Crooks~\cite{sivak_2012}. 

The paper proceeds as follows: \S\ref{sec:background} introduces the relevant theoretical background information; \S\ref{sec:infinite_work} derives an exact expression for the energetic cost of a series of discrete steps applied to an equilibrium system, and then investigates a small-perturbation approximation of this expression to compare to established results; \S\ref{sec:noneq_work} extends this analysis to systems which are out of equilibrium, quantifying within the linear-response regime the excess energetic cost due to incomplete relaxation; \S\ref{sec:optimal_protocols} discusses general implications of optimal, minimum-work protocols in discretely driven systems, under the assumptions laid out in the previous sections; finally, \S\ref{sec:harmonic} and \S\ref{sec:periodic} explore the quantitative implications of this theoretical framework in two model systems: a translating harmonic trap and a periodic potential.

\section{\label{sec:background}Background}

We consider a system in contact with a heat bath, subject to a set of experimentally controlled parameters $\bsla$, such that at equilibrium the distribution over microstates is
\begin{equation}
\pi(x|\bsla) = e^{-\beta E(x,\bsla) + \beta F(\bsla)} \label{boltzmann} \ ,
\end{equation}
where $\beta \equiv (\kT)^{-1}$ is the inverse temperature, $E(x,\bsla)$ is the system Hamiltonian, and $F(\bsla)$ is the equilibrium free energy at control parameter vector $\bsla$. 

A \emph{control protocol} $\Lambda:\bsla_{0}\to\bsla_{N}$ is a particular time-dependent perturbation applied to the control parameter vector $\bsla$ to transform it between an initial $\bsla_{0}$ and final $\bsla_{N}$ in a prescribed time $\tau$. For a given control protocol, the system responds stochastically. Across the entire control protocol $\Lambda$, the average amount of \emph{excess work} (supplied by an external source)---or work required above and beyond the equilibrium free energy difference $\beta \Delta F_{\rm tot}$ between the initial and final control parameter values---is 
$\langle \beta W_{\rm ex}\rangle_{\Lambda} \equiv \langle \beta W\rangle_{\Lambda} - \beta \Delta F_{\rm tot}$, where $\langle\cdots\rangle_{\Lambda}$ indicates an average of system responses to the control protocol $\Lambda$.

Here, we consider discrete control protocols which consist of a series of instantaneous perturbations $\Delta\bsla_{i,i+1} \equiv \bsla_{i+1} - \bsla_i$ for consecutive control parameter values $\bsla_i$ and $\bsla_{i+1}$. The system spends a prescribed time $\ti$ at each control parameter $\bsla_i$. Thus, each protocol is defined by a set of control parameter values and the associated times spent at them:
\begin{equation}
\Lambda \equiv \{ \lambda_i, \ti \} \label{discrete_protocol} \ .
\end{equation}
	
Previous work has considered energetic flows in discrete-stepping processes~\cite{nulton_1985,andreson_1984,salamon_1983,nulton_1984}, but these efforts typically focused on the continuum limit. More recent investigations of driven nonequilibrium systems~\cite{sivak_2012,schmiedl_2007,zulkowski_2012,zulkowski_2014,zulkowski_2015,sivak_2016} have focused on control protocols which are continuous functions of time. In contrast to these previous works, and motivated by the chemically driven paradigm characteristic of microscopic machines, we consider a control protocol as a series of discrete steps of substantial size. 

The average work (divided by $k_{\rm B}T$) associated with a particular discrete perturbation that transforms the control parameter from $\bsla_i$ to $\bsla_{i+1}$ during the protocol $\Lambda$ is 
\begin{equation}
\langle \beta W \rangle_{\bsla_i \to \bsla_{i+1}}
= \beta \int\left[ E(x,\bsla_{i+1}) - E(x,\bsla_i) \right]p_{\Lambda}(x,t_{i,i+1}) \, \md x  \label{work_discrete_background} \ ,   
\end{equation}
where $p_{\Lambda}(x,t_{i,i+1})$ is the (generally nonequilibrium) distribution over system microstates $x$ at the time $t_{i,i+1}$ that the control parameter changes from $\bsla_i$ to $\bsla_{i+1}$ during the protocol $\Lambda$, and angle brackets $\langle\cdots\rangle_{\bsla_i \to \bsla_{i+1}}$ indicate a nonequilibrium average as the control parameter changes from $\bsla_i$ to $\bsla_{i+1}$ during protocol $\Lambda$.
    
We consider protocols which start in equilibrium at initial control parameter value $\bsla_{0}$, equivalent to taking the time spent at $\bsla_{0}$ to infinity, $\tz\to\infty$. A particular control protocol begins with the first control parameter change $\bsla_0\to\bsla_1$, and finishes when the control parameter $\bsla$ arrives at its terminal value $\bsla_{N}$. The protocol duration $\tau$ is the time taken to complete a given protocol, not counting the time taken to equilibrate at the initial control parameter value. Thus, for a protocol with $N+1$ control parameter values $\bsla_0,\bsla_1,\ldots,\bsla_N$, the protocol duration is
\begin{equation}
\tau \equiv \sum_{i=1}^{N-1}\ti \label{protocol_duration} \ ,
\end{equation}
and the total control parameter displacement for fixed control parameter endpoints $\bsla_0,\bsla_N$ is
\begin{equation}
\Delta\bsla_{\rm tot} \equiv \sum_{i=0}^{N-1} \Delta\bsla_{i,i+1} 
= \bsla_N - \bsla_0 
\label{CP_distance} \ .
\end{equation}
For given control parameter endpoints $\bsla_0$ and $\bsla_N$ (hence given protocol displacement $\Delta\bsla_{\rm tot}$) and duration $\tau$, minimizing work involves choosing intermediate control parameter values $\bsla_i$ and associated dwell times $\ti$.

\section{\label{sec:infinite_work}Infinite-time work}
We first consider the work associated with making a single discrete change to the control parameter of a system which is initially at equilibrium with the control parameter $\boldsymbol{\lambda}_{0}$~\eref{boltzmann}. The average work required to discretely change the control parameter vector from $\boldsymbol{\lambda}_0$ to $\boldsymbol{\lambda}_1$ is \eref{work_discrete_background}, with the equilibrium initial distribution $ p_{\Lambda}(x,t_{0,1}) = \pi (x|\bsla_0)$~\eref{boltzmann}:
\begin{equation}
\langle \beta W\rangle_{\bsla_0\to \bsla_1} = \beta \int \left[ E(x,\bsla_1) - E(x,\bsla_0) \right]\pi(x|\bsla_0)\md x \label{work_general} \ .
\end{equation}
From the definition of the equilibrium ensemble at a fixed control parameter~\eref{boltzmann}, the energy can be expressed in terms of the equilibrium distribution and free energy, $\beta E(x,\bsla_i) = -\ln \pi(x|\bsla_i) + \beta F(\bsla_i)$. The average work from \eref{work_general} can then be written solely in terms of equilibrium distributions,
\begin{equation}
\langle \beta W\rangle_{\bsla_0\to \bsla_1} = \int \ln\left[ \frac{\pi(x|\bsla_0)}{\pi(x|\bsla_1)} \right] \pi(x|\bsla_0)\md x + 
\beta \Delta F_{0,1}
\label{work_log_distribution} \ ,
\end{equation}
where $\beta\Delta F_{0,1} = \beta F(\bsla_1) - \beta F(\bsla_0)$ is the difference between the equilibrium free energy at control parameter $\bsla_0$ and $\bsla_1$. The integral in \eref{work_log_distribution} is the \emph{relative entropy} (or Kullback-Leibler divergence) $D[p(x)||q(x)] \equiv \int \ln[p(x)/q(x)]\, p(x)\md x$~\cite{cover_thomas} between the equilibrium distributions before ($p(x) = \pi(x|\bsla_0)$) and after ($q(x) = \pi(x|\bsla_1)$) the control parameter change,
\begin{equation}
\langle \beta W \rangle_{\bsla_0\to \bsla_1} =  D[\pi(x|\bsla_0) \, \| \, \pi(x|\bsla_1)] + \beta\Delta F_{0,1} \label{work_relative_entropy} \ .
\end{equation}

A protocol consists of $N$ such control parameter steps $\Delta\bsla_{i,i+1}$, for $i=0, \ldots, N-1$. If at each control parameter value $\bsla_i$ the system fully equilibrates, then the average work associated with \emph{any} step $\Delta\bsla_{i,i+1}$ is of the same form~\eref{work_relative_entropy}, and the work to complete the entire protocol is the sum	of the work associated with each individual step,
\begin{equation}
\langle \beta W \rangle_{\Lambda} = \sum_{i=0}^{N-1}D\left[\pi(x||\bsla_{i}) \, \| \, \pi(x|\bsla_{i+1})\right] + \beta\Delta F_{\rm tot} \label{protocol_work_inifinite_time} \ ,
\end{equation} 
where $\beta\Delta F_{\rm tot} \equiv \beta F(\bsla_N) - \beta F(\bsla_0) = \sum_{i=0}^{N-1} \beta \Delta F_{i,i+1}$ is the equilibrium free energy change between the initial and final control parameter values. Thus the average excess work is
\numparts
\begin{eqnarray}
\langle \beta W_{\rm ex}\rangle_{\Lambda} &\equiv \langle \beta W\rangle_{\Lambda} - \beta\Delta F_{\rm tot} \\
&= \sum_{i=0}^{N-1} D[\pi(x|\bsla_{i}) \, \| \, \pi(x|\bsla_{i+1})] \label{protocol_excess_work_infinite_limit} \ .
\end{eqnarray}
\endnumparts
	
For sufficiently small control parameter steps $\Delta\bsla_{i,i+1}$, the relative entropy in \eref{protocol_excess_work_infinite_limit} can be Taylor expanded about its current value $\bsla_i$ to yield
\begin{equation}
D\left[ \pi(x|\bsla_i) \, \| \, \pi(x|\bsla_{i+1}) \right] \approx \frac{1}{2}\beta^2\langle\delta f_j\delta f_k\rangle_{\bsla_i}\Delta\lambda_{i,i+1}^j\Delta\lambda_{i,i+1}^k\label{relative_entropy_fisher_approximation} \ ,
\end{equation}
where $\langle\delta f_j\delta f_k\rangle_{\bsla_i}$ is the equilibrium covariance of conjugate forces $f_j \equiv -\partial_{\lambda^j}E$ at control parameter $\bsla_i$ (see \ref{appendix:relative_entropy} for details)~\cite{sivak_2012}. Throughout, we employ the Einstein summation notation, where repeated indices are implicitly summed over. 

Substituting \eref{relative_entropy_fisher_approximation} in \eref{protocol_excess_work_infinite_limit} gives the average excess work required to perform the infinite-time discrete protocol $\Lambda$,
\begin{equation}
\langle \beta W_{\rm ex}\rangle_{\Lambda} \approx \frac{1}{2}\beta^2\sum_{i=0}^{N-1} \langle \delta f_j\delta f_k\rangle_{\bsla_i} \Delta\lambda_{i,i+1}^j\Delta\lambda_{i,i+1}^k \ , \label{protocol_work_fisher}
\end{equation}
which in the continuous-protocol limit is equivalent to the Burbea-Rao entropy differential metric~\cite{burbea_1982} and the thermodynamic metric derived by Crooks~\cite{crooks_2007}.

\section{\label{sec:noneq_work}Nonequilibrium excess work}
To consider the more general situation of finite-time protocols, where at each control parameter value the system does not fully equilibrate, we appeal to static linear-response theory~\cite{chandler}. For a system at equilibrium for control parameter $\bsla_{i-1}$, the energy at the next control parameter value $\bsla_{i}$ in the protocol $\Lambda$ can be linearly approximated as
\numparts
\begin{eqnarray}
E(x,\bsla_i) & \approx E(x,\bsla_{i-1}) + \nabla_{\bsla} E(x,\bsla)|_{\bsla_{i-1}} \cdot \Delta\bsla_{i-1,i} \label{linear-perturbation-1} \\
&= E(x,\bsla_{i-1}) - \boldsymbol{f}|_{\bsla_{i-1}} \cdot  \Delta\bsla_{i-1,i} \label{linear-perturbation-2} 
\end{eqnarray}
\endnumparts
where $\boldsymbol{f}|_{\bsla_{i-1}}$
is the vector of conjugate forces with elements $f_j$, evaluated at $\bsla_{i-1}$.  When the control parameter instantaneously changes from $\bsla_{i-1}$ to $\bsla_i$, the time-dependent relaxation of $f_j$ towards its equilibrium value at $\bsla_{i}$ is, under the linear-response approximation,
\begin{equation}
\langle f_j(\tgen)\rangle_{\bsla_{i-1},\bsla_i} = \langle f_j\rangle_{\bsla_{i}} + \beta\langle\delta f_j(0)\delta f_k(\tgen)\rangle_{\bsla_{i}}\Delta\lambda_{i-1,i}^k \label{static_linear_response_step} \ , 
\end{equation}
where $\langle f_j(\tgen) \rangle_{\bsla_{i-1},\bsla_i}$ indicates an average over the instantaneous \emph{nonequilibrium} system distribution after relaxing (under the Hamiltonian $E(x,\bsla_i)$) for a time $\tgen$ starting from the equilibrium distribution at $\bsla_{i-1}$. Both the force autocovariance and the average force on the RHS are taken over the \emph{equilibrium} ensemble at fixed control parameter value $\boldsymbol{\lambda}_{i}$. \ref{appendix:static_linear_response} provides a detailed derivation of~\eref{static_linear_response_step}.
The second RHS term in \eref{static_linear_response_step} is a linear-response relaxation function, which have a long history of usage to understand the dynamic properties of nonequilibrium systems, for instance in chemical relaxation kinetics~\cite{bernasconi} and chemical dynamics~\cite{Chandler:1978hh,Skinner:1978ef,Kutz:1974em}.

If the system relaxes for a time $\ti$ at control parameter $\boldsymbol{\lambda}_{i}$ before the next control parameter change $\bsla_{i}\to \bsla_{i+1}$, this step requires average work~\eref{work_discrete_background}
\numparts
\begin{eqnarray}
\langle \beta W\rangle_{\bsla_i \to \bsla_{i+1}} &= \beta\Delta\lambda^j_{i,i+1}\langle f_j(\ti)\rangle_{\bsla_{i-1} , \bsla_i} \label{work_general_conjugate_force_vector} \\ 
&\approx \beta\Delta\lambda_{i,i+1}^j\langle f_j\rangle_{\bsla_{i}}
+ \beta^2\Delta\lambda_{i,i+1}^j\langle\delta f_j(0)\delta f_k(\ti)\rangle_{\bsla_{i}}\Delta\lambda_{i-1,i}^k  \label{work_general_conjugate_force_element} \\ &= \langle \beta W^{\infty}\rangle_{\bsla_i\to \bsla_{i+1}} + \langle \beta W_{\rm ex}^{\rm neq}\rangle_{\bsla_i\to \bsla_{i+1}} \label{work_general_conjugate_force_final}
\end{eqnarray}\label{work_general_conjugate_force}
\endnumparts
for	infinite-time work $\langle \beta W^{\infty}\rangle = \beta \Delta\lambda_{i,i+1}^j\langle f_j\rangle_{\bsla_i}$ from \S\ref{sec:infinite_work} and linear-response correction $\langle \beta W_{\rm ex}^{\rm neq}\rangle$ due to incomplete system relaxation.

\eref{work_general_conjugate_force_final} is only strictly valid if the system was at equilibrium for $\bsla_{i-1}$ before the step to $\bsla_{i}$, so more generally the work required for the control parameter change $\bsla_{i}\to\bsla_{i+1}$ includes contributions from all previous steps. However, \eref{work_general_conjugate_force_final} approximates the work when the force autocovariance associated with the most recent step is the largest time-dependent contribution to the excess work. This limit is reached when the time spent at each control parameter value is long compared to the relaxation time of the conjugate forces. As discussed in detail in \ref{appendix:dynamic_linear_response}, this approximation is fundamentally distinct from approximations made in deriving the continuous-protocol formalism~\cite{sivak_2012}.

Within this limit, the total average excess work required to perform a discrete control protocol $\Lambda$ is
\numparts
\begin{eqnarray}
\fl \langle \beta &W_{\rm ex}\rangle_{\Lambda} \equiv \sum_{i=0}^{N-1} \langle \beta W_{\rm ex} \rangle_{\bsla_i \to \bsla_{i+1}} \\
\fl &\approx \langle \beta W_{\rm ex}^{\infty}\rangle_{\Lambda} + \beta^2\sum_{i=1}^{N-1}\Delta\lambda_{i,i+1}^j\langle\delta f_j(0)\delta f_k(\ti)\rangle_{\bsla_{i}}\Delta\lambda_{i-1,i}^k \label{full_work_1}  \\
\fl &\approx \beta^2 \sum_{i=0}^{N-1}\frac{1}{2}\langle\delta f_j\delta f_k\rangle_{\bsla_i}\Delta\lambda_{i,i+1}^j\Delta\lambda_{i,i+1}^k + \beta^2\sum_{i=1}^{N-1}\Delta\lambda_{i,i+1}^j\langle\delta f_j(0)\delta f_k(\ti)\rangle_{\bsla_i}\Delta\lambda_{i-1,i}^k \label{full_excess_work_2} \ ,
\end{eqnarray}\label{full_excess_work_1_2}
\endnumparts
where \eref{full_excess_work_2} uses the infinite-time excess work for small steps~\eref{protocol_work_fisher}. The assumption that the system begins in equilibrium at $\bsla_0$ ensures that the $i=0$ term does not contribute to the total nonequilibrium excess work. By setting $\bsla_{-1} = \bsla_0$, both sums in \eref{full_excess_work_1_2} can be taken over the same index range, leading to a more compact form for the excess work:
\begin{equation}
\fl \langle \beta W_{\rm ex}\rangle_{\Lambda} = \beta^2\sum_{i=0}^{N-1}\langle\delta f_j\delta f_k\rangle_{\bsla_i} \Delta\lambda_{i,i+1}^j\Delta\lambda_{i,i+1}^k \left[ \frac{1}{2} + \frac{\langle\delta f_j(0)\delta f_k(\ti)\rangle_{\bsla_i}}{\langle\delta f_j\delta f_k\rangle_{\bsla_i}} \frac{\Delta\lambda_{i-1,i}^k}{\Delta\lambda_{i,i+1}^k} \right] \label{cost_function} 
\ .
\end{equation}
This captures the combined effects of the control parameter step sizes $\Delta\bsla_{i,i+1}$ and time allocations $\ti$ on the excess work during a discrete control protocol $\Lambda$. The time-independent, infinite-time contribution penalizes large control parameter steps departing from regions with large force covariance. The time-dependent linear-response correction penalizes steps that are particularly quick (reflected by the force autocorrelation factor) and/or large (reflected by the $\Delta\lambda_{i,i+1}^j\Delta\lambda_{i-1,i}^k$ step-size factor). 

\Eref{cost_function} generalizes the near-equilibrium expression for the dissipation of a discrete control protocol in \cite{nulton_1985}, because the time dependence captured by the conjugate-force autocovariance in our approach allows for non-exponential relaxation kinetics, and the explicit form permits simultaneous optimization of both the placement of control parameter values $\bsla_i$ as well as the allocation of times $\ti$. Moreover, \ref{appendix:dynamic_linear_response} provides an alternative derivation of the nonequilibrium excess work contribution~\eref{work_general_conjugate_force_final} using dynamic linear response theory to show that in the continuous-protocol limit, a  linear-response correction to the excess work can recover the generalized friction formalism from \cite{sivak_2012}, but only if approximations are made which are incompatible with those used to derive~\eref{cost_function}. 

\section{\label{sec:optimal_protocols}Minimum-work protocols}
The nonequilibrium excess work~\eref{cost_function} provides a relatively simple expression, within the linear-response approximation, for the energetic cost required to perform discrete control protocol $\Lambda$. Although the specific form of a minimum-work protocol depends on the particular system, there are two special cases admitting analytic solutions: the case of the infinite-time limit (\S\ref{sec:infinite_work}) where the time-dependent term in \eref{cost_function} is negligible, and the case where there is a single dominant exponential relaxation mode.
    
In general, the excess work can be approximated as
\numparts
\begin{eqnarray}
\langle \beta W_{\rm ex}\rangle_{\Lambda} &= \sum_{i=0}^{N-1} \langle \beta W_{\rm ex} \rangle_{\bsla_i \to \bsla_{i+1}} \\
&\equiv \sum_{i=0}^{N-1} \Gamma_{jk}(\bsla_{i-1},\bsla_{i},\bsla_{i+1},\ti)\Delta\lambda_{i,i+1}^j\Delta\lambda_{i,i+1}^k \label{cost_function_metric_1} \\
&\equiv \sum_{i=0}^{N-1}\mathcal{D}_i^2 \label{cost_function_metric_2} \ ,
\end{eqnarray}\label{cost_function_metric}
\endnumparts
and 
\begin{equation}
\fl \Gamma_{jk}(\bsla_{i-1},\bsla_{i},\bsla_{i+1},\ti) \equiv \beta^2\langle\delta f_j\delta f_k\rangle_{\bsla_i} \left[ \frac{1}{2} + \frac{\langle \delta f_j(0)\delta f_{k}(\ti)\rangle_{\bsla_i}}{\langle\delta f_j\delta f_{k} \rangle_{\bsla_i}} \frac{\Delta\lambda_{i-1,i}^{{k}}}{\Delta\lambda_{i,i+1}^{{k}}} \right] \ , \nonumber
\end{equation}
where, in this case, the $k$ index within the brackets is not summed over. Interpretation is most immediate in the continuous-protocol limit, where each $\mcD_i$ is the distance along an infinitesimal segment $\md\bsla$ of the control protocol $\Lambda$, measured with respect to the metric $\Gamma_{jk}(\bsla_{i-1},\bsla_i,\bsla_{i+1},\ti)$; therefore, the sum $\sum_{i=0}^{N}\mcD_i$ over all steps gives the \emph{thermodynamic length} between the initial and final equilibrium system macrostates~\cite{crooks_2007,nulton_1985}. 
    
For a positive semidefinite force-autocovariance matrix, the total excess work of a particular control protocol can be lower-bounded via the Cauchy-Schwarz inequality:
\begin{equation}
\langle \beta W_{\rm ex}\rangle_{\Lambda} \geq \frac{1}{N}\left(\sum_{i=0}^{N}\mathcal{D}_i\right)^2 \label{work-lower-bound} \ .
\end{equation}
The lower bound is saturated if and only if the $\mcD_i$ are identical, 
\begin{equation}
    \mcD_i = \mcD \label{optimal-condition} \ .
\end{equation}
Along an optimal protocol (indicated by the superscript $*$), the condition~\eref{optimal-condition} implies that $\langle \beta W_{\rm ex}^{*}\rangle_{\bsla_i\rightarrow \bsla_{i+1}} = \mcD^2$, and thus equal excess work is done during each step of the protocol.

For a single control parameter with fixed endpoints $\lambda_0,\lambda_N$ and a given set of time allocations $\ti$, the condition~\eref{optimal-condition} implies the optimal placement of control parameter values through the proportionality 
\begin{equation}
    \Delta\lambda_{i,i+1}^* \propto \frac{1}{\sqrt{\Gamma(\lambda_{i-1}^*,\lambda_{i}^*,\lambda_{i+1}^*,\ti)}} \label{CS-bound} \ ,
\end{equation}
but the implicit dependence of $\Gamma$ on the step size through $\lambda_{i-1}^*,\lambda_{i}^*$, and $\lambda_{i+1}^*$ complicates the practical use of this bound for deriving optimal protocols. However, the proportionality~\eref{CS-bound} can give useful qualitative guidance into the general properties of protocols which saturate the lower bound~\eref{work-lower-bound}. In particular, optimal control parameter placement tends to avoid regions with large force variance and slowly decaying force autocovariance, subject to the quadratic cost $\Delta\lambda_{i,i+1}^j\Delta\lambda_{i,i+1}^k$ on step sizes. 
For more than one control parameter, the qualitative insights gained from the lower bound~\eref{work-lower-bound} and the equality~\eref{optimal-condition} can provide a way to derive the optimal time-schedule along a particular path in control parameter space, but unfortunately they do not generally provide a constructive means to identify a path that saturates the bound.

In the infinite-time limit, where $\Gamma_{jk}(\bsla_{i-1},\bsla_{i},\bsla_{i+1},\ti)\to \Gamma_{jk}^{(\infty)}(\bsla_i) = \frac{1}{2}\beta^2\langle\delta f_j\delta f_k\rangle_{\bsla_i}$, our predictions reduce to previous calculations by Nulton \textit{et al.}~\cite{nulton_1985} of the optimal placement of discrete steps. In particular, for a single control parameter, the condition~\eref{optimal-condition} implies that optimal protocols have the proportionality 
$\Delta\lambda_{i,i+1}^* \propto 1/\sqrt{\langle\delta f^2\rangle_{\lambda_i^*}}$. Furthermore, in the continuous-protocol limit, the infinite-time thermodynamic length between the initial and final control parameters (measured with respect to $\Gamma_{jk}^{(\infty)}(\bsla_i)$) converges to that of Crooks~\cite{crooks_2007}. 

The optimal time allocation is analytically solvable for a single control parameter when the time dependence in \eref{cost_function} takes a simple, control-parameter-dependent, exponential form: 
\begin{equation}
\frac{\langle\delta f(0)\delta f(\ti)\rangle_{\lambda_i}}{\langle\delta f^2\rangle_{\lambda_i}} = e^{-\ti/\tau_{\rm R}(\lambda_i)} \ .
\end{equation}
Using Lagrange multipliers, the optimal allocation of time among a fixed set of control parameter values, subject to the protocol duration constraint~\eref{protocol_duration}, is
\begin{equation}
\ti^* = \tau_{\rm R}(\lambda_i)\left[\frac{\tau}{\sum_{n=1}^{N-1}\tau_{\rm R}(\lambda_{n})} - 
\frac{\sum_{n=1}^{N-1}\tau_{\rm R}(\lambda_{n})\ln\left(\mathcal{P}_{n}/\mathcal{P}_i\right)}{\sum_{{n}=1}^{N-1}\tau_{\rm R}(\lambda_{n})}
\right]  \label{optimal-time-allocation} \ ,
\end{equation}
where $\mathcal{P}_{n} \equiv \beta^2\Delta\lambda_{n,n+1}\Delta\lambda_{n-1,n}\langle\delta f^2\rangle_{\lambda_{n}}/\tau_{\rm R}(\lambda_{n})$ (see \ref{appendix:protocol_optimization} for a detailed derivation). 

In the long-duration limit, where $\tau \gg \sum_{s=1}^{N-1}\tau_{\rm R}(\lambda_{s})$, the second RHS term in \eref{optimal-time-allocation} is negligible, and the optimal allocation of time takes on the simple form
\begin{equation}
\ti^* \propto \tau_{\rm R}(\lambda_i) \label{optimal-time-allocation-2} \ .
\end{equation} 
Intuitively, this implies that along minimum-work protocols, more time is allocated to regions where the integral relaxation time~\cite{garanin_1996} is larger.
    
The special case of a single control parameter and exponential relaxation kinetics produces \eref{optimal-time-allocation}, which recovers the result of Nulton \textit{et al.} in \cite{nulton_1985}.
However, our more general framework~\eref{cost_function} can be applied to a broader class of problems (as we detail in \S\ref{sec:harmonic} and \S\ref{sec:periodic}), in particular to cases with multiple control parameters, non-exponential relaxation kinetics, and optimization of control parameter placements. For more general scenarios, even with one control parameter, analytic optimization methods become cumbersome, and no simple analogs of~\eref{optimal-time-allocation} can be found.  Nevertheless, \eref{cost_function} provides a relatively simple expression that can be minimized using numerical methods (\ref{appendix:protocol_optimization} provides more details).

\section{\label{sec:harmonic}Harmonic trap}

We now focus on a system diffusing in a one-dimensional harmonic trap defined by the potential
\begin{equation}
E_{\rm t}(x,\lambda_{i}) = \frac{1}{2}k_{\rm t}(x - \lambda_i)^2 \label{harmonic_potential} \ .
\end{equation}
Here $k_{\rm t}$ is the trap strength and the control parameter is the time-dependent trap minimum $\lambda_i$. The work required to perform an $N$-step discrete control protocol $\Lambda$, taking the control parameter from its initial value $\lambda_0$ to $\lambda_N$, can be calculated exactly. 

\subsection{\label{subsec:harmonic_trap}Infinite-time limit}

In the infinite-time limit (\S\ref{sec:infinite_work}), the excess work for a single step $\lambda_i\to\lambda_{i+1}$ is the relative entropy~\eref{work_relative_entropy} between the equilibrium distributions~\eref{boltzmann} at $\lambda_i$ and $\lambda_{i+1}$:
\begin{eqnarray}
\langle \beta W_{\rm ex}\rangle_{\lambda_{i}\to\lambda_{i+1}} &= D\left[\pi(x|\lambda_{i})||\pi(x|\lambda_{i+1})\right]\nonumber \\ &= \frac{1}{2} \beta k_{\rm t} \Delta\lambda_{i,i+1}^2 \label{harmonic_work_infinite_step} \ .
\end{eqnarray}
Thus the infinite-time work for a discrete protocol of $N$ steps is
\begin{equation}
\langle \beta W_{\rm ex}\rangle_{\Lambda} = \frac{1}{2} \beta k_{\rm t} \sum_{i=0}^{N-1}\Delta\lambda_{i,i+1}^2 \label{harmonic_infinite_work_protocol} \ .
\end{equation}
Based on the convexity of this expression, equal step sizes $\Delta\lambda_{i,i+1} = \Delta\lambda_{\rm tot}/N$ minimize the infinite-time work,
\begin{equation}
\langle \beta W_{\rm ex}\rangle_{\Lambda} \geq \frac{1}{2} \beta k_{\rm t} \sum_{i=0}^{N-1}\left( \frac{\Delta\lambda_{\rm tot}}{N} \right)^2 = \frac{\beta k_{\rm t} \Delta\lambda_{\rm tot}^2}{2N} \label{harmonic_infinite_work_bound} \ ,
\end{equation} 
which scales with the number of steps as $1/N$~\cite{burbea_1982}. For this simple system~\eref{harmonic_potential}, the small-step approximation of the relative entropy~\eref{protocol_work_fisher} is exact, for arbitrary step sizes.
	
\subsection{\label{subsec:harmonic_trap_general}General solution: finite-time work}

Finite-duration control protocols feature both the infinite-time excess work and the time-dependent contribution (\S\ref{sec:noneq_work}). For a system initially in equilibrium at the initial control parameter $\lambda_0$, the average excess work (in this case equal to the total work since $\Delta F = 0$) for the $(i+1)$th step is
\begin{equation}
\langle \beta W_{\rm ex}\rangle_{\lambda_i \to \lambda_{i+1}} = \beta k_{\rm t}\Delta\lambda_{i,i+1}^2\left[ \frac{1}{2} + \frac{\xi_{i-1} e^{-\beta D k_{\rm t}\ti}}{\Delta\lambda_{i,i+1}} \right] \label{harmonic_work_step_i} \ ,
\end{equation}
where 
\numparts
\begin{eqnarray}
\xi_{i-1} &\equiv \sum_{n=0}^{i-1}\Delta\lambda_{n,n+1}\exp\left( -\beta D k_{\rm t} \sum_{r=n+1}^{i-1} \tgen_{r} \right) \label{xi_term_1} \\
&= \Delta\lambda_{i-1,i} + \Delta\lambda_{i-2,i-1}e^{- \beta D k_{\rm t}\tgen_{i-1}} + \cdots \label{xi_term_2} \ .
\end{eqnarray}\label{xi_term}
\endnumparts
($\xi_{-1} = 0$ as there are no terms in that summation.) \ref{appendix:harmonic_trap} provides a detailed derivation of (\ref{harmonic_work_step_i},\ref{xi_term_1}).

Summing \eref{harmonic_work_step_i} over the entire protocol $\Lambda$ gives
\begin{equation}
\langle \beta W_{\rm ex}\rangle_{\Lambda} = \beta \sum_{i=0}^{N-1} k_{\rm t}\Delta\lambda_{i,i+1}^2\left( \frac{1}{2} + \frac{\xi_{i-1} e^{-\beta D k_{\rm t}\ti}}{\Delta\lambda_{i,i+1}} \right) \label{harmonic_work_protocol_exact} \ .
\end{equation}
For this simple system, the normalized force autocovariance (the force autocorrelation) is $\langle \delta f(0)\delta f(\ti)\rangle_{\lambda_i}/\langle\delta f^2\rangle_{\lambda_i} = \exp(-\beta D k_{\rm t} \ti)$, so the approximate excess work within the linear-response regime~\eref{cost_function} is
\begin{equation}
\langle \beta W_{\rm ex}\rangle_{\Lambda} = \beta \sum_{i=0}^{N-1}k_{\rm t}\Delta\lambda_{i,i+1}^2\left( \frac{1}{2} + \frac{\Delta\lambda_{i-1,i}}{\Delta\lambda_{i,i+1}} e^{-\beta D k_{\rm t} \ti}\right) \label{harmonic_work_protocol_LR} \ ,
\end{equation}	
which is equivalent to truncating $\xi_{i-1}$ from \eref{xi_term_2} after the first term: $\xi_{i-1} \approx \Delta\lambda_{i-1,i}$. From \eref{xi_term_1}, it follows that the linear-response approximation~\eref{cost_function} holds when
\begin{equation}
\frac{\Delta\lambda_{i,i+1}}{\Delta\lambda_{i-1,i}} \gg e^{-\beta D k_{\rm t} \ti} \label{harmonic_approx_condition} \ ,
\end{equation}
which is satisfied in the limit of long times $\ti \gg 1/(\beta D k_{\rm t})$ spent at each control parameter. 
	
For a protocol $\Lambda$ of protocol duration $\tau$ and consisting of $N$ control parameter steps, each of uniform size $\Delta\lambda_{i,i+1} = \Delta\lambda_{\rm tot}/N$, with uniform time allocations $\ti = \tau/(N-1) \equiv \tgen_{\rm step}$, the exact excess work is
\begin{equation}
\langle \beta W_{\rm ex}\rangle_{\Lambda} = \frac{\beta k_{\rm t}\Delta\lambda_{\rm tot}^2}{N^2}\sum_{i=0}^{N-1}\left( \frac{1}{2} + \frac{\xi_{i-1}}{\Delta\lambda_{i,i+1}} e^{-\beta D k_{\rm t}\tgen_{\rm step}} \right) \ , \label{harmonic_naive_exact_work}
\end{equation}
the infinite-time work is \eref{harmonic_infinite_work_protocol}, and for large $N$ the linear-response work is
\begin{equation} 
\langle \beta W_{\rm ex} \rangle_{\Lambda} \approx \frac{\beta k_{\rm t}\Delta\lambda_{\rm tot}^2}{N}\left( \frac{1}{2} + e^{-\beta D k_{\rm t}\Delta t_{\rm step}} \right)
\label{harmonic_naive_approx_work} \ ,
\end{equation} 

\begin{figure}[ht]
	\centering\includegraphics[width=\textwidth]{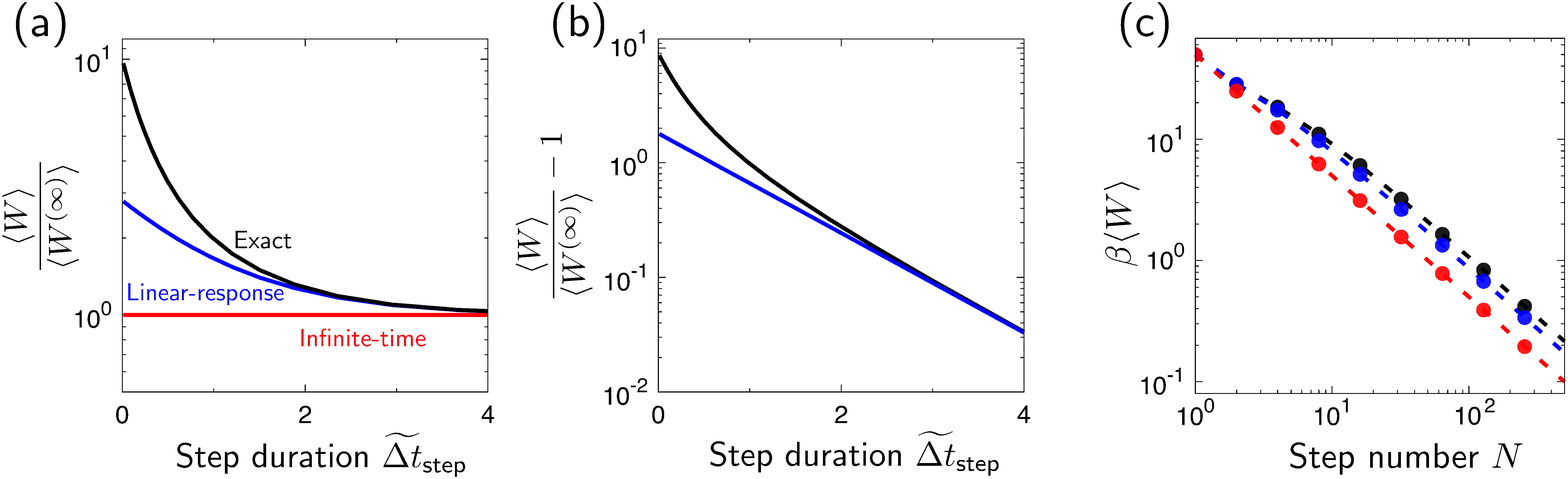}
	\caption{\label{fig:harmonic_exact}{\bf Exact and approximate work for a discretely driven harmonic trap.} 
	Black: exact solution~\eref{harmonic_naive_exact_work}; blue: linear-response approximation~\eref{harmonic_naive_approx_work}; red: infinite-time limit~\eref{harmonic_infinite_work_bound}. (a) Excess work, normalized by the infinite-time limit~\eref{harmonic_infinite_work_protocol}, as a function of the nondimensionalized step duration $\tND_{\rm step} \equiv \beta D k_{\rm t} \tau/(N-1)$ (scaled by the number of relaxation times spent at each control parameter value). (b) The difference between the normalized excess work and its infinite-time limit of unity scales exponentially for longer step durations ($\tND_{\rm step} \gtrapprox 2$), and also converges to the linear-response prediction. (c) Average protocol work $\beta\langle W\rangle$ for a fixed step duration $\tND_{\rm step} = 1$, as a function of the number $N$ of control parameter steps. The predicted $1/N$ scaling is seen in the exact solution and linear-response approximation at sufficiently large $N$, and in the infinite-time limit at all $N$. All plots are for uniform step spacing, $\Delta\lambda_{i,i+1} = \Delta\lambda_{\rm tot}/N$ for each step $i$.}
\end{figure}

In each case, for a fixed duration $\tgen_{\rm step}$ allocated to each control parameter value, the work scales asymptotically ($N\to\infty$) as $1/N$. Figure~\ref{fig:harmonic_exact}a and b show the average excess work for $N=10$ and $\Delta\lambda_{\rm tot}/N = 1$, and the difference between the average work and the infinite-time limit as a function of the step duration. For sufficiently large step duration, the exact result~\eref{harmonic_naive_exact_work} converges to the linear-response prediction~\eref{harmonic_naive_approx_work} and exponentially approaches the infinite-time limit. The three curves have a fixed ordering: the exact solution~\eref{harmonic_naive_exact_work} has a series of positive terms added beyond the linear-response expression~\eref{harmonic_naive_approx_work}, which in turn has an extra positive term added beyond the infinite-time limit~\eref{harmonic_infinite_work_bound}.

For protocol durations sufficiently long that the time spent at each control parameter significantly exceeds the relaxation time, the linear-response approximation and the exact result converge. 
Furthermore, neglecting the $\xi_{i-1}$ term reduces the exact protocol work~\eref{harmonic_work_protocol_exact} to the infinite-time limit~\eref{harmonic_infinite_work_protocol}. Figure~\ref{fig:harmonic_exact}c shows, for the particular step-duration $\tgen_{\rm step} = (\beta D k_{\rm t})^{-1}$, the $1/N$ scaling (for large $N$) of the average protocol work.

\section{\label{sec:periodic}Periodic potential}
\begin{figure}
    \centering\includegraphics[width=\textwidth]{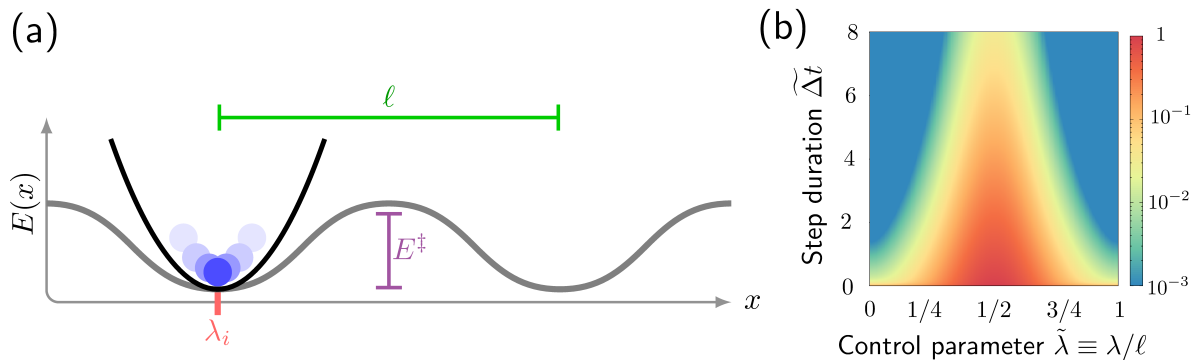}
    \caption{{\bf (a) Schematic depiction of the periodic potential.} The control parameter $\lambda_i$ (red) takes discrete steps $\Delta\lambda_{i,i+1}$ to drive the system (fluctuating blue ball) over a series of energy barriers (of height $E^{\ddagger}$) separating adjacent metastable potential wells. The underlying potential has period $\ell$ (green). {\bf (b) Force autocovariance sampled across a single period of the underlying potential.} Heat map for scaled force autocovariance $\langle \delta f(0)\delta f(\tgen)\rangle_{\lambda}/\max_{\lambda}(\langle \delta f^2\rangle_{\lambda})$ as a function of control parameter $\tilde{\lambda} \equiv \lambda/\ell$. $\tND \equiv \tgen/\overline{\tau_{\rm relax}}$ is the nondimensionalized step duration, and $\overline{\tau_{\rm relax}}$ is the period-averaged integral relaxation time.}
	\label{fig:periodic_schematic_combo}
\end{figure}

Now we consider a single diffusing particle in a one-dimensional energy landscape $E(x,\lambda_i)$ consisting of two components: a control-parameter-independent periodic potential, and a control-parameter-dependent harmonic trapping potential:
\begin{equation}
E_{\rm periodic}(x,\lambda_i) = \frac{1}{2} k_{\rm t} \left( x - \lambda_i \right)^2 - \frac{1}{2}E^{\ddagger} \cos \frac{2\pi x}{\ell} \label{periodic_potential} \ ,
\end{equation}
where $k_{\rm t}$ is the harmonic trap strength, $E^{\ddagger}$ is the energetic barrier height between adjacent minima on the periodic potential, and $\ell$ is the period (Fig.~\ref{fig:periodic_schematic_combo}). This potential represents a system with a sequence of metastable states, such as those often found in models of molecular machines~\cite{reimann_2002}.
	
Figure~\ref{fig:periodic_schematic_combo}b shows numerical estimation of the autocovariance $\langle\delta f(0)\delta f(\tgen)\rangle_{\lambda}$ from equilibrium simulations at several fixed control parameters evenly spaced over a single period of the underlying potential~\eref{periodic_potential}. Using the force autocovariance as input to the linear-response approximation~\eref{cost_function}, we minimize the average excess work during a discrete control protocol $\Lambda$ with a fixed number $N$ of steps. \ref{appendix:simulation_details} gives details on the equilibrium simulations and numerical optimization of the excess work.

We consider three different protocol optimization schemes in order to isolate the effects of the optimal allocation of times $\ti$ to a fixed `naive' sequence of control parameter values (a `time-optimized' protocol), the optimal placement of control parameters for a fixed `naive' set of time allocations (`space-optimized'), and the simultaneous optimization of time allocations and control parameter placements (`fully optimized'). In all cases, protocols are constrained by having fixed protocol duration $\tau$~\eref{protocol_duration}, protocol endpoints $\lambda_0,\lambda_N$, and number of steps $N$. In order to minimize the effect of the boundary conditions, we consider control protocols which traverse several periodic repetitions of the underlying potential. 
	
\begin{figure}[ht]
    \centering\includegraphics[width=0.8\textwidth]{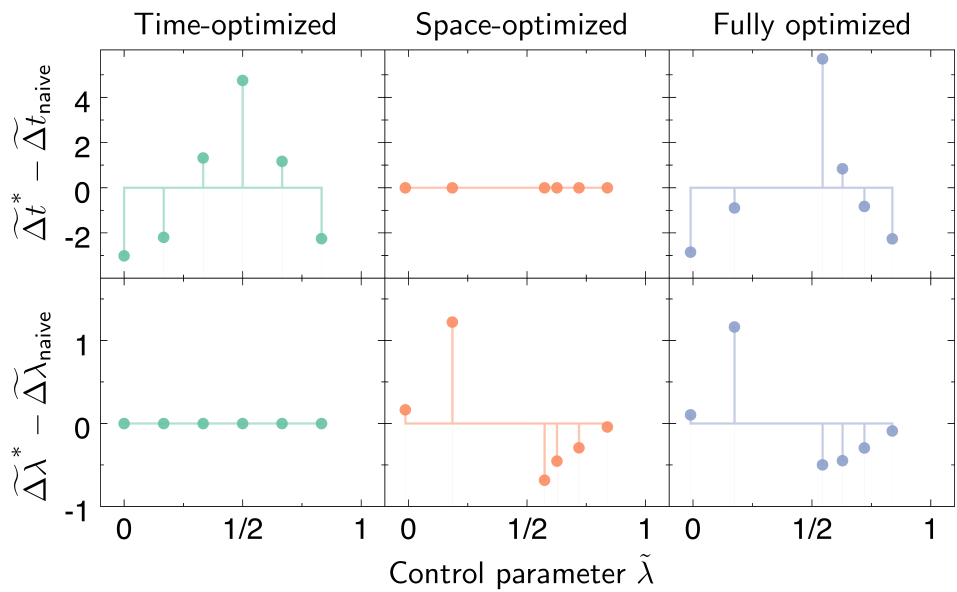}
    \caption{\label{fig:protocol-figure}{
    \bf Protocols designed to minimize work allocate time and/or steps significantly differently from naive protocols.} 
    Relative (nondimensionalized) time allocations $\tND^* - \tND_{\rm naive}$ (top row) and relative step sizes $\widetilde{\Delta\lambda}^* - \widetilde{\Delta\lambda}_{\rm naive}$ (bottom row), for time-optimized (left column), space-optimized (middle column), and fully optimized discrete protocols (right column), with $N_{\rm p}=6$ steps per periodic repetition of the underlying landscape. $\tgen_{\rm naive} \equiv \tau/(N-1)$ is the naive time allocation, and $\Delta\lambda_{\rm naive} \equiv \Delta\lambda_{\rm tot}/N$ is the naive control parameter step size. Time allocations are nondimensionalized as $\tND \equiv \tgen/(N\overline{\tau_{\rm relax}})$, where $\overline{\tau_{\rm relax}}$ is the mean integral relaxation time over a single period of the underlying potential. Control parameter step sizes are nondimensionalized as $\widetilde{\Delta\lambda} \equiv \Delta\lambda/(N_{\rm p}\ell)$, where $N_{\rm p}$ is the number of steps in a periodic repetition of the underlying potential, and $\ell$ is the period of the underlying potential. Protocol has a total duration $\tau = 4(N-1)\overline{\tau_{\rm relax}}$.} 
\end{figure}

For such a discrete protocol, Fig.~\ref{fig:protocol-figure} shows the time allocations $\ti$ and control parameter step-sizes $\Delta\lambda_{i,i+1}$, relative to their naive values $\tgen_{\rm naive} \equiv \tau/(N-1)$ and $\Delta\lambda_{\rm naive} \equiv \Delta\lambda_{\rm tot}/N$, as a function of the control parameter value $\lambda$ over a single period, for a protocol with $N_{\rm p} = 6$ steps per potential period and protocol duration $\tau = 4(N-1)\overline{\tau_{\rm relax}}$, where $\overline{\tau_{\rm relax}}$ is the mean integral relaxation time over a single period of the underlying potential.

In each case, the behavior predicted by our theoretical analysis of simplified systems in \S\ref{sec:optimal_protocols} is borne out. In particular, time-optimized protocols allocate a larger fraction of the protocol duration to regions where the force is slowly relaxing~\eref{optimal-time-allocation-2}, while space-optimized protocols take step sizes which are largest in regions where the force variance is small and rapidly relaxing. In the fully optimized protocols, both behaviors are present. 

\begin{figure}[ht]
\centering\includegraphics[width=0.8\textwidth]{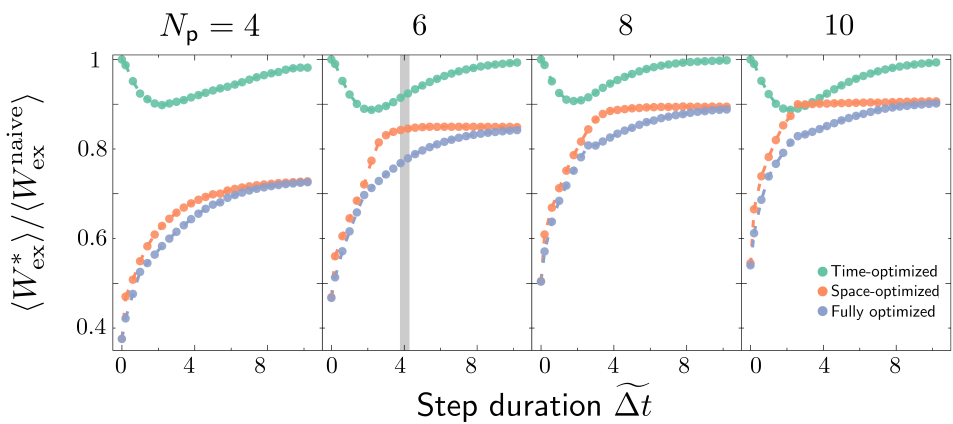}
\caption{\label{fig:discrete_control_results}{\bf Optimized discrete control protocols significantly reduce the predicted excess work}. Excess work ratio for discrete control protocols traversing several periodic images, as a function of nondimensionalized step duration $\tND \equiv \tau/(N\overline{\tau_{\rm relax}})$. Purple: fully optimized protocols~\eref{cost_function}; turquoise: time-optimized; orange: space-optimized. Number $N_{\rm p}$ of steps per potential period varies from left to right sub-plots. The grey bar on the $N_{\rm p} = 6$ subplot indicates the protocols shown in Fig.~\ref{fig:protocol-figure}.}
\end{figure}

Figure~\ref{fig:discrete_control_results} shows the theoretically expected excess work for these minimum-work protocols, specifically the predicted excess work ratio $\langle W_{\rm ex}^{*}\rangle_{\Lambda}/\langle W_{\rm ex}^{\rm naive}\rangle_{\Lambda}$ for the three distinct protocol classes: time-optimized, space-optimized, and fully optimized. For short durations $(\tND \equiv \tau/(N\overline{\tau_{\rm relax}}) < 1)$, time optimization yields no gain over naive protocols, while spatial optimization and full optimization are indistinguishable. For intermediate durations ($\tND \approx2$), time optimization has maximum effect, and full optimization significantly improves upon spatial optimization. For longer durations ($\tND \gtrapprox 8$), time optimization again gives no benefit over the naive protocol, as the time-dependent term in \eref{cost_function} becomes negligible.
   
\begin{figure}
\centering\includegraphics[width=0.9\textwidth]{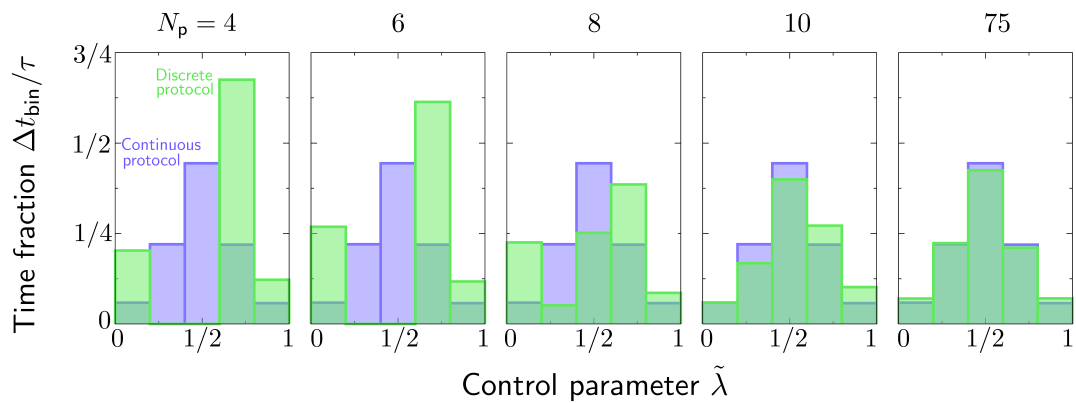}
\caption{{\bf In the many-step/continuous-protocol limit, fully optimized discrete protocols allocate time the same as their continuous-protocol analogs.} 
The fraction of the total protocol duration spent in each region of control parameter space for a fully optimized discrete control protocol (green), compared to the fraction during an optimized continuous protocol generated using the generalized friction framework~\cite{sivak_2012}. As the number $N_{\rm p}$ of control parameter values per potential period increases from $N_{\rm p} = 4$ to $N_{\rm p} = 75$, the discrete-protocol histogram converges to the continuous-protocol histogram.}
\label{fig:continuous-limit}
\end{figure}

Figure~\ref{fig:continuous-limit} shows that as the number $N_{\rm p}$ of steps per periodic image increases, the time allocation for fully optimized discrete control protocols converges to that of the optimal continuous protocol derived from the generalized friction coefficient~\cite{sivak_2012}.

However, there is a significant difference between the discrete at low step numbers $N_{\rm p}$ and continuous control protocols~\cite{sivak_2012,schmiedl_2007}. In particular, relative to an optimal continuous protocol, fully optimized discrete protocols allocate a smaller fraction of their duration at $\lambda = \ell/2$, near the energy barrier (and in fact at the lowest step numbers completely avoid this region). This ability of low step-number protocols to entirely avoid regions of control parameter space with high force variance and slow relaxation (generally speaking, near energy barriers) represents a qualitatively distinct optimization strategy that is simply not available to continuous protocols.
    
Furthermore, the continuous protocols allocate time symmetrically about the energy barrier because the generalized friction maintains the same symmetries as the underlying energetic landscape~\eref{periodic_potential}~\cite{sivak_2012}. As a result, a continuous optimal protocol traverses the same path in both the forward $\lambda_0\to\lambda_{\rm N}$ and reverse $\lambda_{\rm N}\to \lambda_0$ directions. Discrete protocols break this symmetry because of the infinite-time contribution (\S\ref{sec:infinite_work}) as, in general, $D[p(x)||q(x)] \neq D[q(x)||p(x)]$. For small steps~\eref{protocol_work_fisher}, this asymmetry persists; in \eref{cost_function} the excess work during the control parameter step $\Delta\lambda_{i,i+1}$ is a function of the force variance at the \emph{current} control parameter $\langle\delta f^2\rangle_{\lambda_i}$ (and independent of the force variance $\langle\delta f^2\rangle_{\lambda_{i+1}}$ at the destination control-parameter value $\lambda_{i+1}$). This produces a directional asymmetry as the excess work for the control parameter step $\Delta\lambda_{i+1,i}$ is generally different than the excess work for step $\Delta\lambda_{i,i+1}$.  However, as the number of control parameter steps increases and the distance between those steps becomes sufficiently small, the difference between the force variance at consecutive control parameter values becomes negligibly small, and the asymmetry between forward and reverse protocols vanishes (Fig.~\ref{fig:continuous-limit}).
    
\section{\label{sec:discussion}Discussion}
	
In this article, we derived the work required to drive a microscopic system out of equilibrium via a discrete control protocol. Such a control protocol transforms the energy landscape through a series of discrete intermediate states, capturing the discrete nature of the chemical reaction sequences that drive many biological molecular motors. The central result is the linear-response expression for excess work~\eref{cost_function}, which quantifies the near-equilibrium work of a particular control protocol, as a function solely of the equilibrium system properties.
	
We deduced a general expression for the work required to make a discrete change in the control parameter vector of a system in equilibrium~\eref{protocol_excess_work_infinite_limit} and used this to exactly quantify the work required to perform a discrete protocol in the infinite-time limit (\S\ref{sec:infinite_work}). When each step is sufficiently small and hence each perturbation is sufficiently weak, our derivation reduces to previously known results~\cite{burbea_1982,crooks_2007}. Our primary contribution is to generalize these analyses outside of the infinite-time limit, where we use a linear-response approximation to derive the leading-order time-dependent contribution to the excess work (\S\ref{sec:noneq_work}). Theoretically, this work goes significantly beyond previous efforts~\cite{nulton_1985} to quantify energy flows in discretely driven nonequilibrium systems, by incorporating the effects of relaxation kinetics that are non-exponential and that vary across control parameter space, and by simultaneously optimizing both the placement of control parameter values as well as the allocation of times. 
	
We investigated the correspondence between our linear-response approximation and an exact solution for a harmonically trapped Brownian particle driven by a series of discrete steps of equal size. We also studied the optimal allocation of time and placement of control parameter values that minimize the work for protocols traversing many repetitions of a periodic energy landscape~\eref{periodic_potential}. We find that fully optimized discrete control protocols have qualitatively distinct features when compared to their continuous-protocol analogs. In particular, discrete protocols do not obey the same directional symmetry that continuous protocols do, and in the context of the periodic potential (\S\ref{sec:periodic}), discrete protocols allocate a smaller fraction of their total duration near the energy barrier. More generally, minimum-work protocols allocate more time to regions where the force has a smaller variance and is more slowly decaying. Finally, we quantified the reduction in excess work, relative to a naive protocol, achieved by these minimum-work discrete control protocols. In particular, the theoretical excess work reduction (relative to a naive protocol) of a fully optimized protocol exceeds $50\%$ for small step numbers and short protocol durations.  Significant reduction persists even for intermediate durations ($\tilde{\tau}\approx 2$) when fully optimized, highlighting the benefits of both optimized placement of control parameter values and the allocation of time among them.
		
The paradigm of a discretely driven nonequilibrium system is motivated, in part, due to its resemblance to the chemical driving in many biomolecular machines. Stochastic protocol ensembles have been recently considered in related work~\cite{large_2018_EPL}, and future investigation on how the stochastic properties of chemical driving affect the dissipation in discretely driven systems promises a more robust framework within which to compare the theoretical predictions for optimal operation to experimental results on the natural operation of molecular machines.

\section{\label{sec:acknowledgements}Acknowledgements}
    
The authors thank John Bechhoefer and Emma Lathouwers (SFU Physics) and Miranda Louwerse (SFU Chemistry) for insightful comments on the article.
This work is supported by Natural Sciences and Engineering Research Council of Canada (NSERC) CGS Masters and Doctoral Fellowships (S.J.L.), an NSERC Discovery Grant (D.A.S.), a Tier-II Canada Research Chair (D.A.S.), and WestGrid (www.westgrid.ca) and Compute Canada Calcul Canada (www.computecanada.ca).

\section*{References}

\providecommand{\newblock}{}

\appendix

\section{\label{appendix:relative_entropy}Expansion of the relative entropy}
The relative entropy (Kullback-Leibler divergence) between two continuous probability distributions $p(x)$ and $q(x)$ is defined as~\cite{cover_thomas}
\begin{equation}
D[q(x)||p(x)] \equiv \int \ln\left[ \frac{p(x)}{q(x)} \right] p(x)\md x \ . \label{appendix_relative_entropy}
\end{equation}
In the context of the present work, the equilibrium distribution $\pi(x|\bsla_i)$ is parameterized by the control variable $\bsla$. The integrand of the relative entropy for two consecutive equilibrium distributions at $\bsla_i$ and $\bsla_{i+1}$ is
\begin{equation}
g(x,\bsla_i,\bsla_{i+1}) = \pi(x|\bsla_i) \ln \frac{\pi(x|\bsla_i)}{\pi(x|\bsla_{i+1})} \label{perturbation_relative_entropy_integrand}.
\end{equation} 
For small changes $\Delta\bsla_{i,i+1} \equiv \bsla_{i+1} - \bsla_i$ in the control parameter, we Taylor expand Eq.~\eref{perturbation_relative_entropy_integrand} about $\bsla_i$,
\begin{eqnarray}
\fl g(x,\bsla_{i},\bsla_{i+1}) &= g(x,\bsla_i,\bsla_{i}) + \left[\partial_{\lambda_{i+1}^j}g(x,\bsla_{i},\bsla_{i+1})\right]_{\lambda_i^j}\Delta\lambda_{i,i+1}^j \label{appendix_integrand_expansion_1} \\
&\quad + \frac{1}{2}\left[\partial_{\lambda_{i+1}^j}\partial_{\lambda_{i+1}^k}g(x,\bsla_i,\bsla_{i+1})\right]_{\lambda_i^j,\lambda_i^k}\Delta\lambda_{i,i+1}^j\Delta\lambda_{i,i+1}^k + \mathcal{O}(\Delta\bsla^3) \nonumber \ , 
\end{eqnarray}
where $\partial_{\lambda_{i+1}^m}g(x,\bsla_i,\bsla_{i+1}) \equiv \frac{\partial}{\partial\lambda_{i+1}^m}g(x,\bsla_i,\bsla_{i+1})$ is the partial derivative of $g(x,\bsla_i,\bsla_{i+1})$ with respect to the $m$th component of the control parameter $\bsla_{i+1}$, $\left[\cdots\right]_{\lambda_i^m}$ indicates that the argument is evaluated at $\lambda_{i+1}^m = \lambda_i^m$, and we have made use of the Einstein summation notation, where repeated indices are summed over.

The first term in \eref{appendix_integrand_expansion_1} is
\begin{equation}
g(x,\bsla_i,\bsla_{i}) = \pi(x|\bsla_i) \ln\frac{\pi(x|\bsla_i)}{\pi(x|\bsla_i)} = \pi(x|\bsla_i) \ln 1 = 0 \ .
\end{equation}
The derivative on the RHS of~\eref{appendix_integrand_expansion_1}, evaluated at $\bsla_{i+1} = \bsla_i$ is
\begin{equation}
\partial_{\lambda^j_{i+1}}\left\{ \pi(x|\bsla_i) \ln \frac{\pi(x|\bsla_i)}{\pi(x|\bsla_{i+1})} \right\}_{\lambda_i^j} = -\frac{\partial \pi(x|\bsla_{i+1})}{\partial\lambda_{i+1}^j}\bigg|_{\lambda^j_i} \label{expansion_term_1}\ ,
\end{equation}
and the second-derivative term in \eref{appendix_integrand_expansion_1} is
\begin{eqnarray}
\fl \partial_{\lambda_{i+1}^j}\partial_{\lambda_{i+1}^k}\left\{ \pi(x|\bsla_i) \ln \frac{\pi(x|\bsla_i)}{\pi(x|\bsla_{i+1})} \right\}_{\lambda_i^j,\lambda_i^k} &= \frac{1}{\pi(x|\bsla_i)}\left[ \frac{\partial \pi(x|\bsla_{i+1})}{\partial\lambda_{i+1}^j} \frac{\partial \pi(x|\bsla_{i+1})}{\partial\lambda_{i+1}^k}\right]_{\lambda_i^j,\lambda_i^k} \nonumber \\
&\quad - \frac{\partial^2 \pi(x|\bsla_{i+1})}{\partial\lambda_{i+1}^j\partial\lambda_{i+1}^k}\bigg|_{\lambda_i^j,\lambda_i^k} \label{expansion_term_2} 
\ .
\end{eqnarray}

Equation~\eref{expansion_term_1} can be simplified by noting that the equilibrium probability distribution is normalized, $\int \pi(x,\bsla)\md x = 1$, and partial differentiation commutes with integration, so substituting \eref{expansion_term_1} into the relative entropy expression~\eref{appendix_relative_entropy}, gives
\begin{equation}
\frac{\partial}{\partial\lambda_{i+1}^j} \int \pi(x|\bsla_{i+1}) \, \md x = \frac{\partial}{\partial\lambda_{i+1}^j}1 = 0 \ , \label{normalization_zero_term_1}
\end{equation}
so this term does not contribute to the overall relative entropy. This results from the relative entropy being a convex function with a minimum at $\Delta\bsla = 0$. In analogy with \eref{normalization_zero_term_1}, the second term on the RHS of \eref{expansion_term_2} also vanishes,
\begin{equation}
\frac{\partial^2}{\partial\lambda_{i+1}^k\partial\lambda_{i+1}^j}\int\pi(x|\bsla_{i+1})\, \md x = \frac{\partial^2}{\partial\lambda_{i+1}^k\partial\lambda_{i+1}^j}1 = 0 \ .
\label{normalization_zero_term_2}
\end{equation}
Combining \eref{normalization_zero_term_1} and \eref{normalization_zero_term_2} with \eref{appendix_integrand_expansion_1}, the relative entropy for a control parameter step $\Delta\bsla_{i,i+1}$ is
\begin{eqnarray} 
\fl D\left[\pi(x|\bsla_i)||\pi(x|\bsla_{i+1})\right] \label{appendix_relative_entropy_final} \\
= \frac{1}{2} \Delta\lambda_{i,i+1}^j\Delta\lambda_{i,i+1}^k \int \frac{1}{\pi(x|\bsla_i)}\left[ \frac{\partial\pi(x|\bsla_{i+1})}{\partial\lambda_{i+1}^j}\frac{\partial \pi(x|\bsla_{i+1})}{\partial\lambda_{i+1}^k} \right]_{\bsla_i}{\rm d}x + \mathcal{O}(\Delta\bsla^3) \nonumber \ ,
\end{eqnarray}
where the integral is the Fisher information matrix $\mathcal{I}_{jk}(\bsla_i)$ at control parameter $\bsla_i$~\cite{cover_thomas}. For sufficiently small steps, the $\mathcal{O}(\Delta\bsla^3)$ term is negligible, so for a discrete control protocol $\Lambda$, consisting of $N$ steps, the excess work in the infinite-time limit
\eref{protocol_excess_work_infinite_limit} is
\begin{equation}
\langle \beta W_{\rm ex}\rangle_{\Lambda} = \frac{1}{2}\sum_{i=0}^{N-1}\mathcal{I}_{jk}(\bsla_i)\Delta\lambda_{i,i+1}^j\Delta\lambda_{i,i+1}^k \label{excess_work_protocol_fisher} \ .
\end{equation}

For a physical system in contact with a thermal reservoir, the equilibrium distribution is \eref{boltzmann}. Within the linear-response regime the energy can be expanded about $\bsla_0$~\eref{linear-perturbation-2},
\begin{equation}
E(x,\bsla) \approx E(x,\bsla_0) - f_j|_{\bsla_0}(\lambda^j - \lambda_0^j) + \mathcal{O}(\Delta\bsla^2)
\end{equation}
for the conjugate force $f_j|_{\bsla_0} \equiv -\partial E/\partial\lambda^j|_{\bsla_0}$ (evaluated at $\bsla_0$). Derivatives of the equilibrium distribution are
\begin{equation}
\partial_{\lambda^j}\pi(x|\bsla) = \beta\left( f_j|_{\bsla} + \frac{\partial F(\bsla)}{\partial\lambda^j} \right) \pi(x|\bsla) \ . \label{appendix_partial_eq_dist}
\end{equation}
From the thermodynamic definition of the free energy,
\begin{eqnarray}
F(\bsla) &= \langle E\rangle_{\bsla} - T S \\
&= -\lambda^j\langle f_j\rangle_{\bsla} - T S \ ,
\end{eqnarray}
so partial derivatives of the free energy in \eref{appendix_partial_eq_dist} are $\partial_{\lambda^j}F(\bsla) = -\langle f_j\rangle_{\bsla}$. Therefore, for an equilibrium distribution~\eref{boltzmann}, the Fisher information is
\begin{eqnarray}
\mathcal{I}_{jk}(\bsla_i) &= \int\frac{1}{\pi(x|\bsla_i)}\left[ \partial_{\lambda^j}\pi(x|\bsla)\partial_{\lambda^k}\pi(x|\bsla) \right]_{\bsla_i} \\
&=\beta^2\int \left( f_j|_{\bsla_i} - \langle f_j\rangle_{\bsla_i} \right)\left( f_k|_{\bsla_k} - \langle f_k\rangle_{\bsla_i} \right)\pi(x|\bsla_i)\, \md x	\\
&= \beta^2 \langle\delta f_j\delta f_k\rangle_{\bsla_i} \label{fisher_to_variance} \ ,
\end{eqnarray}
where $\delta f_j \equiv f_j|_{\bsla_i} - \langle f_j\rangle_{\bsla_i}$. Substituting \eref{fisher_to_variance} into \eref{excess_work_protocol_fisher} gives
\begin{equation}
\langle \beta W_{\rm ex}\rangle_{\Lambda} = \frac{1}{2}\beta^2\sum_{i=0}^{N-1}\langle\delta f_j\delta f_k\rangle_{\bsla_i}\Delta\lambda_{i,i+1}^j\Delta\lambda_{i,i+1}^k \label{excess_work_protocol_variance} \ .	
\end{equation}
This final equation is equivalent to the infinite-time protocol work~\eref{protocol_work_fisher} in \S\ref{sec:infinite_work}.

\section{\label{appendix:static_linear_response}Nonequilibrium excess work: static linear response}
We consider a system with control parameter vector $\bsla_i$ which is in contact with a thermal reservoir, so that the equilibrium distribution over microstates is \eref{boltzmann}. Within the linear-response regime, the Hamiltonian for
$\bsla_{i-1}$ is
\begin{equation}
E(x,\bsla_{i-1}) \approx E(x,\bsla_i) - f_j|_{\bsla_{i}} \Delta\lambda_{i-1,i}^j  \ .
\end{equation}
For a system initially (at $t = 0$) at equilibrium for $\bsla_{i-1}$,
\begin{eqnarray}
p(x|\bsla_{i-1},t=0) &= e^{-\beta E(x,\bsla_{i-1}) + \beta F(\bsla_{i-1})} \\ 
&= e^{-\beta(E(x,\bsla_{i}) - f_j\Delta\lambda_{i-1,i}^j) + \beta F(\bsla_{i-1})} \label{appendix_static_LR_initial_dist} \ .
\end{eqnarray}
At $t = t_{i-1,i}$, the control parameter instantaneously switches from $\bsla_{i-1}\to\bsla_i$, such that initially the distribution over microstates~\eref{appendix_static_LR_initial_dist} is a nonequilibrium distribution evolving under the Hamiltonian $E(x,\bsla_{i})$. According to linear-response theory, the time-dependent average of the difference of the $j$th element of the conjugate force vector $\boldsymbol{f}$ from its equilibrium value at $\bsla_i$, as a function of the time $\tgen$ passed since the Hamiltonian was instantaneously perturbed, is~\cite{chandler}
\begin{equation}
\langle\delta f_j(\tgen)\rangle_{\bsla_{i-1},\bsla_i} = \beta\langle\delta f_j(0)\delta f_k(\tgen)\rangle_{\bsla_i}\Delta\lambda_{i-1,i}^k \label{appendix_static_LR_neq_avg} \ .
\end{equation}
Here $\delta f_j \equiv f_j|_{\bsla_i} - \langle f_j\rangle_{\bsla_i}$ denotes an instantaneous deviation of the conjugate force from its equilibrium value at $\bsla_i$, angle brackets $\langle\cdots\rangle_{\bsla_{i-1},\bsla_i}$ indicate an average over the instantaneous \emph{nonequilibrium} distribution of a system relaxing from an equilibrium state at $\bsla_{i-1}$ towards the equilibrium at $\bsla_i$, whereas angle brackets $\langle\cdots\rangle_{\bsla_i}$ indicate an average over equilibrium fluctuations at fixed control parameter $\bsla_i$.
	
The work required to change the control parameter $\bsla_i\to\bsla_{i+1}$ after spending a time $\ti$ at $\bsla_i$ is
\begin{equation}
\langle W \rangle_{\bsla_i\to\bsla_{i+1}} = \langle f_j(\ti)\rangle_{\bsla_{i-i},\bsla_i} \Delta\lambda_{i,i+1}^j \ . \label{appendix_static_LR_work}
\end{equation}
Substituting \eref{appendix_static_LR_neq_avg} into \eref{appendix_static_LR_work}, the average work during the control parameter change $\bsla_i\to\bsla_{i+1}$, given that the system was previously at equilibrium with $\bsla_{i-1}$ at $\tgen = 0$, and has since spent a time $t_i$ relaxing towards equilibrium at $\bsla_i$, is
\begin{eqnarray}
\langle W\rangle_{\bsla_i \to \bsla_{i+1}} &= \langle f_j(\ti)\rangle_{\bsla_{i-1},\bsla_i} \Delta\lambda_{i,i+1}^j \\
&= \langle f_j\rangle_{\bsla_i} \Delta\lambda_{i,i+1}^j + \langle\delta f_j(\ti)\rangle_{\bsla_{i-1},\bsla_i} \Delta\lambda_{i,i+1}^j \\ 
&= \langle W_{\infty}\rangle_{\bsla_i\to \bsla_{i+1}} + \beta\langle\delta f_j(0)\delta f_k(\ti)\rangle_{\bsla_i} \Delta\lambda_{i-1,i}^k \Delta\lambda_{i,i+1}^j \ .
\label{appendix_static_LR_final}
\end{eqnarray}		
In the third line, we identified $\langle f_j\rangle_{\bsla_i} \Delta\lambda_{i,i+1}^j$ as the work required to perturb the system from $\bsla_i$, given that it has equilibrated there, i.e., the infinite-time work discussed in \S\ref{sec:infinite_work} in the main text.

\section{\label{appendix:dynamic_linear_response}Nonequilibrium excess work: dynamic linear response}
We have considered a system in contact with a thermal reservoir and subject to the control parameter vector $\bsla$, so that the equilibrium distribution over microstates is \eref{boltzmann}. When subjected to a control protocol $\Lambda$, dynamic linear-response theory says that at a time $t'$ after the start of the protocol, the average deviation $\langle\delta f_j(t')\rangle_{\Lambda}$ of the $j$th component of the conjugate force from its equilibrium value at the current control parameter is~\cite{sivak_2012}
\begin{equation}
\langle \delta f_j(t')\rangle_{\Lambda}
\approx \int_{-\infty}^{t'}\frac{\md}{\md t''}\left[ \langle\delta f_j(0)\delta f_k(t'-t'')\rangle_{\bsla(t')} \right]\lambda^k(t'') \md t'' \label{appendix_LR_dynamic} \ .
\end{equation}
Here $\langle\delta f_j(t')\rangle_{\Lambda}$ indicates an average of the conjugate force fluctuation at time $t'$ over system response subject to the protocol $\Lambda$, and the integral ranges over the entire previous history of the control protocol $\Lambda$. 
	
For a discrete control protocol, the time-dependent history $\lambda^k(t'')$ can be represented by a sum of weighted Heaviside functions 
\begin{equation}
\Delta\lambda_{i,i+1}^k\theta(t'' - t_{i,i+1}) \equiv 
\cases{
0 \ , &$\quad t''\leq t_{i,i+1}$ \cr
\Delta\lambda^k_{i,i+1} \ , &$\quad t'' > t_{i,i+1}$ \cr
}
\label{appendix:heaviside}
\end{equation}
reflecting the control parameter jumps of size $\Delta\bsla_{i,i+1}$, occurring at time $t_{i,i+1}$. Using \eref{appendix:heaviside}, the $j$th component of the force fluctuation~\eref{appendix_LR_dynamic} becomes
\begin{equation}
\fl \langle\delta f_j(t')\rangle_{\Lambda} \approx \int_{-\infty}^{t'}\frac{\md}{\md t''}\left[ \langle\delta f_j(0) \delta f_k(t'-t'')\rangle_{\bsla_i} \right] \sum_{n=0}^{i-1}\Delta\lambda_{n,n+1}^k\theta(t'' - t_{n,n+1}) \md t'' \label{appendix_LR_dynamic_2}\ ,
\end{equation}
where $\bsla_i$ is the current control parameter value, and the average $\langle\delta f_j(t')\rangle_{\Lambda}$ accounts for the contributions due to all previous steps in the discrete protocol $\Lambda$.

Integrating \eref{appendix_LR_dynamic_2} by parts, observing that the boundary term is zero if the system begins in thermodynamic equilibrium, and substituting the Dirac delta function for the derivative of the Heaviside function, $\partial_{t''}\theta(t'' - t_{n,n+1}) = \delta(t'' - t_{n,n+1})$, gives
\begin{equation}
\langle\delta f_j(t')\rangle_{\Lambda}	\approx \int_{-\infty}^{t'}\langle\delta f_j(0)\delta f_k(t' - t'')\rangle_{\bsla_i} \sum_{n=0}^{i-1}\Delta\lambda_{n,n+1}\delta(t'' - t_{n,n+1})\md t'' \label{appendix_LR_dynamic_3} \ .
\end{equation}

Written out term by term, \eref{appendix_LR_dynamic_3} takes the form
\begin{eqnarray}
\langle \delta f_j(t')\rangle_{\Lambda}&\approx \langle\delta f_j(0)\delta f_k(t'-t_{i-1,i})\rangle_{\bsla_i}\Delta\lambda_{i-1,i}^k \label{appendix_LR_dynamic_4}\\
&\quad + \langle\delta f_j(0)\delta f_k(t'- t_{i-2,i-1})\rangle_{\bsla_i}\Delta\lambda_{i-2,i-1}^k + \cdots \nonumber \ , 
\end{eqnarray}
which depends on the times of all previous control parameter jumps. In order to reach the result cited in the main text, we simply truncate the expansion after the first (leading) term. In this approximation, the average force fluctuation after spending a time $\ti = t' - t_{i-1,i}$ at control parameter $\bsla_i$ is
\begin{equation}
\langle \delta f_j(t')\rangle_{\Lambda} \approx \langle\delta f_j(0)\delta f_k(\ti)\rangle_{\bsla_i}\Delta\lambda_{i-1,i}^k \label{appendix_LR_dynamic_final} \ .
\end{equation}
	
Now, instead of truncating the series expansion in \eref{appendix_LR_dynamic_4}, we consider the continuous-protocol limit of the entire expansion, where the protocol duration $\tau$ is fixed while both the step sizes and the times spent at each control parameter $\ti \equiv t_{i,i+1} - t_{i-1,i}$ become infinitesimally small, such that $\Delta\lambda^k_{i-1,i}/\ti \to \md\lambda^k/\md t$. Specifically, when spending equal time $\tgen = \tau/(N-1)$ for each step, \eref{appendix_LR_dynamic_4} becomes for $t' = t_{i,i+1}$
\begin{eqnarray}
\langle\delta f_j(t')\rangle_{\Lambda} &\approx \lim_{ \Delta t,\Delta\lambda_i\to 0} \langle\delta f_j(0)\delta f_k(\Delta t)\rangle_{\bsla_i}\Delta\lambda_{i-1,i}^k \\
&\quad + \langle\delta f_j(0)\delta f_k(2 \Delta t)\rangle_{\bsla_i}\Delta\lambda_{i-2,i-1}^k + \cdots \\
&= \lim_{ \Delta t,\Delta\bsla_i\to 0}\sum_{n=1}^{N}\langle\delta f_j(0)\delta f_k(n\ \Delta t)\rangle_{\bsla_i}\frac{\Delta\lambda_{N-n,N-n+1}^k}{\Delta t} \Delta t \\
&= \int_{0}^{\infty}\langle\delta f_j(0)\delta f_k(t)\rangle_{\bsla(t')}\frac{\md \lambda^k(t)}{\md t}\md t \label{appendix_LR_continuous_limit} \ .
\end{eqnarray}
Equation~\eref{appendix_LR_continuous_limit} is the same expression derived by Sivak and Crooks in \cite{sivak_2012}, and was subsequently simplified by Taylor expanding the velocity term in the integrand to zeroth order about its current value,
\begin{eqnarray}
\frac{\md\lambda^k(t')}{\md t} &= \left[\frac{\md\lambda^k}{\md t}\right]_{t'} + \mathcal{O}\left( \frac{\md^2 \lambda^k}{\md t^2} \right) \\
&\approx \left[\frac{\md\lambda^k}{\md t}\right]_{t'} \ .
\end{eqnarray}
This simplifies \eref{appendix_LR_continuous_limit} to
\begin{eqnarray}
\langle\delta f_j(t')\rangle_{\Lambda} &\approx \left(\int_{0}^{\infty}\langle\delta f_j(0)\delta f_k(t'')\rangle_{\bsla(t')}\md t''\right) \left[\frac{\md \lambda^k}{\md t''}\right]_{t'} \\
&= \zeta_{jk}(\bsla(t'))\left[\frac{\md\lambda^k}{\md t''}\right]_{t'} \label{force_fluct_genfriction} \ ,
\end{eqnarray}
where $\zeta_{jk}(\bsla(t'))$ is the generalized friction tensor~\cite{sivak_2012}.

In summary, the derivation of the average excess work for an explicitly discrete control protocol~\eref{cost_function} truncates the dynamic linear-response expression~\eref{appendix_LR_dynamic_2} after the first order, whereas the continuous-protocol result~\eref{force_fluct_genfriction} includes the influence of all previous perturbations, but approximates the previous control parameter velocities by the current value. In essence, this approximation makes similar claims to the discrete truncation in \eref{appendix_LR_dynamic_final}, in that it assumes that the most recent perturbations are the predominant contributors to the excess work. In light of this, the two derivations can be seen as similar approximations which hold in different circumstances: the former for protocols composed of large discrete steps, and the latter for continuous protocols.

\section{\label{appendix:harmonic_trap}Harmonic trap: exact result}
For a system in contact with a thermal bath, subjected to a harmonic confining potential,
\begin{equation}
E_{\rm t}(x,\lambda) = \frac{1}{2}k_{\rm t}\left( x - \lambda \right)^2 \label{appendix_harmonic_trap}\ ,
\end{equation}
and initially at equilibrium, the initial distribution over microstates is
\begin{equation}
\pi(x|\lambda_0) = \sqrt{\frac{\beta k_{\rm t}}{2\pi}} \, e^{-\frac{1}{2}\beta k_{\rm t}\left( x - \lambda_0 \right)^2} \label{appendix_harmonic_initial_dist} \ .
\end{equation}
The protocol work can be calculated exactly when this system is subjected to a discrete control protocol, which takes the trap minimum through a sequence of positions $\lambda_0,\lambda_1,\cdots,\lambda_N$ (with fixed spring constant).
	
The first step taking $\lambda_0\to\lambda_1$ requires average work
\begin{eqnarray}
\langle W\rangle_{\lambda_0 \to \lambda_1} &= \int \left[ E(x,\lambda_1) - E(x,\lambda_0) \right]\pi(x|\lambda_0)\md x \\
&= \frac{1}{2}k_{\rm t}\Delta\lambda_{0,1}^2 \ .
\end{eqnarray}
After the control parameter change, the system is in a nonequilibrium distribution given by the solution to the 1-dimensional Fokker-Planck equation
\begin{eqnarray}
\partial_t &p(x,t|\lambda_1,\lambda_0) = -\beta D k_{\rm t}\partial_{x}\left[ (x - \lambda_1) p(x,t|\lambda_1) \right] + \frac{1}{2}D\partial_{xx}^2p(x,t|\lambda_1)  \label{appendix_harmonic_fokker-planck} \ ,
\end{eqnarray} 
subject to the initial condition $p(x,t=0|\lambda_1,\lambda_0) = \pi(x|\lambda_0)$. ($D$ is the system diffusion coefficient.) The exact solution is known~\cite{gardiner}: after a time $\tgen_1$ spent at $\lambda_1$, the time-dependent probability distribution is
\begin{equation}
p(x,t_1|\lambda_1,\lambda_0) = \sqrt{\frac{\beta k_{\rm t}}{2 \pi}} \exp\left\{-\frac{1}{2}\beta k_{\rm t}\left( x - \lambda_1 + \Delta\lambda_{0,1}e^{-\beta D k_{\rm t}\tgen_1} \right)^2 \right\} \label{appendix_harmonic_first_relaxation} \ ,
\end{equation}
a Gaussian distribution with time-dependent mean $\lambda_1 - \Delta\lambda_{0,1}e^{-\beta D k_{\rm t}\tgen_1}$, which approaches $\lambda_1$ in the infinite-time ($\tgen_1\to\infty$) limit.
	
If, after time $\tgen_1$, the second control parameter step $\lambda_1\to\lambda_2$ takes place, the resulting average work is 
\begin{eqnarray}
\langle W\rangle_{\lambda_1 \to \lambda_2} &= \int \left[ E(x,\lambda_2) - E(x,\lambda_1) \right]p(x,t_1|\lambda_1,\lambda_0)	\md x \\
&= \frac{1}{2}k_{\rm t}\Delta\lambda_{1,2}^2 + k_{\rm t}\Delta\lambda_{1,2}\Delta\lambda_{0,1} e^{-\beta D k_{\rm t}\tgen_1} \label{appendix_harmonic_second_step_work} \ .
\end{eqnarray}

Again, after the control parameter change $\lambda_1\to\lambda_2$, the system is out of equilibrium with probability distribution solving the Fokker-Planck equation~\eref{appendix_harmonic_fokker-planck}, subject to the initial condition $p(x,t=0|\lambda_2,\lambda_1,\lambda_0,\tgen_1) = p(x,\tgen_1|\lambda_1,\lambda_0)$. This leads to the time-dependent system distribution at $\lambda_2$ after a time $\tgen_2$,
\begin{equation}
p(x,\tgen_2|\lambda_2,\lambda_1,\lambda_0,\tgen_1) = \sqrt{\frac{\beta k_{\rm t}}{2 \pi}} \exp\left\{-\frac{\beta k_{\rm t}}{2}\left( x - \lambda_2 + \xi_1e^{-\beta D k_{\rm t}\tgen_2} \right)^2 \right\} \label{appendix_harmonic_second_relaxation}
\end{equation}
for $\xi_1 \equiv \Delta\lambda_{1,2} + \Delta\lambda_{0,1}e^{-\beta D k_{\rm t}\tgen_1}$. This has the same form for the probability distribution as all higher-order steps 
\begin{eqnarray}
p(x,&\tgen_i|\lambda_i,\lambda_{i-1},\cdots,\lambda_0,\tgen_{i-1},\cdots,\tgen_1) \label{appendix_harmonic_general_relaxation} \\
&= \sqrt{\frac{\beta k_{\rm t}}{2 \pi}} \exp\left\{-\frac{1}{2}\beta k_{\rm t}\left( x - \lambda_i + \xi_{i-1}e^{-\beta D k_{\rm t}\ti} \right)\right\} \nonumber 
\end{eqnarray}
for
\begin{equation}
\xi_{i-1} \equiv \sum_{n=0}^{i-1}\Delta\lambda_{n,n+1}\exp\left(-\beta D k_{\rm t}\sum_{m=n+1}^{i-1}\tgen_m\right) \label{appendix_harmonic_xi_general} \ .
\end{equation}
For the time-dependent distribution in \eref{appendix_harmonic_general_relaxation}, the average work during the step $\lambda_i\to\lambda_{i+1}$ is
\begin{equation}
\langle W \rangle_{\lambda_i \to \lambda_{i+1}} = \frac{1}{2}k_{\rm t}\Delta\lambda_{i,i+1}^2 + k_{\rm t}\Delta\lambda_{i,i+1}\xi_{i-1}e^{-\beta D k_{\rm t}\tgen_i} \label{appendix_harmonic_work_step_general} \ .
\end{equation} 
	
From \eref{appendix_harmonic_work_step_general}, the total work during an arbitrary discrete protocol for a harmonic potential is
\begin{equation}
\langle W \rangle_{\Lambda} = \sum_{i=0}^{N-1} k_{\rm t}\Delta\lambda_{i,i+1}^2\left[ \frac{1}{2} + 
\frac{\xi_{i-1}}{\Delta\lambda_{i,i+1}}e^{-\beta D k_{\rm t}\tgen_i} \right] \ .
\end{equation}

\section{\label{appendix:protocol_optimization}Optimization of control protocols}

Here we describe the constrained optimization of the excess work~\eref{cost_function} for a single control parameter,
\begin{equation}
\langle W_{\rm ex}\rangle_{\Lambda} = \beta\sum_{i=0}^{N-1}\Delta\lambda_{i,i+1}^2\langle\delta f^2\rangle_{\lambda_i}\left[ \frac{1}{2} + \frac{\langle\delta f(0)\delta f(\ti)\rangle_{\lambda_i}}{\langle\delta f^2 \rangle_{\lambda_i}} \right] \ .
\end{equation}
We use Lagrange multipliers, following the method from \cite{nulton_1985} on a similar problem. We define the \emph{Lagrange function}
\begin{equation}
\mathcal{L}
= \langle W_{\rm ex}\rangle_{\Lambda} - \epsilon_{\rm s}\left(\sum_{i=1}^{N-1}\ti - \tau\right) - \epsilon_{\tau}\left( \sum_{i=0}^{N-1} \Delta\lambda_{i,i+1} - \Delta\lambda_{\rm tot}\right) \ , \label{appendix_protocol_lagrange_function}
\end{equation}
incorporating fixed protocol endpoints $\lambda_0,\lambda_N$ and spatial~\eref{CP_distance} and temporal~\eref{protocol_duration} constraints defined in the main text, with respective Lagrange multipliers $\epsilon_{\rm s}$ and $\epsilon_{\tau}$.

For the purposes of analytical investigation, we consider naive-space variable-time protocols, where $\Delta\lambda_{i,i+1} = \Delta\lambda_{\rm tot}/{N}$ for all steps with fixed endpoints at $\lambda_0,\lambda_N$, and the time allocations $\ti$ are variable. We find the optimal allocation of times that minimizes the excess work (indicated by superscript $*$) by extremizing the Lagrange function~\eref{appendix_protocol_lagrange_function} with respect to the allocation of times,
\begin{eqnarray}
0 &= \left[\frac{\partial \mathcal{L}}{\partial t_i}\right]_{t_i^*} \\
&= \beta\Delta\lambda_{i,i+1}\Delta\lambda_{i-1,i}\left[\partial_{t_i}\langle\delta f_j(0)\delta f_k(\ti)\rangle_{\bsla_i}\right]_{t_i^*} - \epsilon_{\tau}		\label{appendix_protocol_lagrange_minimization} \ .
\end{eqnarray}
The brackets $\left[ \cdots \right]_{t_i^*}$ indicate that the argument is evaluated at $\ti^*$. In general, this is not analytically tractable, lacking	a functional form for $\langle \delta f(0)\delta f(\ti)\rangle_{\lambda_i}$. 
	
However, if the autocovariance is dominated by a single relaxing mode (or the time intervals $\ti$ are all long enough that the slowest-relaxing mode dominates) then the autocovariance is
\begin{equation}
\langle\delta f(0)\delta f(\ti)\rangle_{\lambda_i} \approx \langle\delta f^2\rangle_{\lambda_i} e^{-\ti/\tau_{\rm R}(\lambda_i)} \label{appendix_protocol_lagrange_derivative} \ ,
\end{equation}
for the characteristic relaxation time $\tau_{\rm R}(\lambda_i)$. Here, the derivative term in \eref{appendix_protocol_lagrange_minimization} is
\begin{equation}
\left[\partial_{t_i}\langle\delta f(0)\delta f(\ti)\rangle_{\lambda_i}\right]_{\ti^*} = -\frac{\langle \delta f^2\rangle_{\lambda_i}}{\tau_{\rm R}(\lambda_i)}e^{-\ti^*/\tau_{\rm R}(\lambda_i)} \ . \label{appendix_protocol_lagrange_derivative_2}
\end{equation}
Substituting this into \eref{appendix_protocol_lagrange_minimization} produces
\begin{equation}
\ln(-\epsilon_{\tau}) = \ln \frac{\beta\Delta\lambda_{i,i+1}\Delta\lambda_{i-1,i}\langle\delta f^2\rangle_{\lambda_i}}{\tau_{\rm R}(\lambda_i)} - \frac{\ti^*}{\tau_{\rm R}(\lambda_i)} \ . \label{appendix_log_multiplier}
\end{equation}
Summing over all steps $i$ in the protocol gives
\begin{equation}
\ln(-\epsilon_{\tau}) = -\left[ \frac{\tau}{\sum_{i=1}^{N-1}\tau_{\rm R}(\lambda_i)} - \frac{\sum_{i=1}^{N-1}\tau_{\rm R}(\lambda_i)\ln \mathcal{P}_i}{\sum_{i=1}^{N-1}\tau_{\rm R}(\lambda_i)}\right] \ , \label{appendix_multiplier_solution}
\end{equation}
where $\mathcal{P}_i \equiv \beta\Delta\lambda_{i,i+1}\Delta\lambda_{i-1,i} \langle\delta f^2\rangle_{\lambda_i}/\tau_{\rm R}(\lambda_i)$, and we have used \eref{protocol_duration} from the main text. Equating \eref{appendix_multiplier_solution} and \eref{appendix_log_multiplier}, the optimal allocation of time that minimizes the excess work is
\begin{equation}
\ti^* = \tau_{\rm R}(\lambda_i)\left[ \frac{\tau}{\sum_{n=1}^{N-1}\tau_{\rm R}(\lambda_n)} - \frac{\sum_{n=1}^{N-1}\tau_{\rm R}(\lambda_n)\ln(\mathcal{P}_n/\mathcal{P}_i)}{\sum_{n=1}^{N-1}\tau_{\rm R}(\lambda_n)} \right] \label{appendix_optimal_time_allocation} \ .
\end{equation} 

This result is equivalent to a similar calculation performed by Nulton \emph{et al.} in \cite{nulton_1985}. Our result significantly extends this previous work, as we give a general expression for the magnitude of $\mathcal{P}_n$ in terms of physical quantities. For long protocol durations, the first term in brackets dominates, and thus the ratio $\ti^*/\tau_{\rm R}(\lambda_i)$ is independent of $i$, implying that an equal number of relaxation times are spent at each control parameter value: longer durations are allocated to control parameter values with longer relaxation times.
	
The analytical optimization of \eref{cost_function} in more complicated scenarios becomes intractible. For instance, even for a single control parameter, using the Lagrange-multiplier method to optimize \eref{cost_function} simultaneously with respect to both the control parameter $\lambda_i$ and the time allocation $\ti$, it is necessary to know partial derivatives of the force autocovariance with respect to both $\lambda_i$ and $\ti$. Furthermore, the optimizations cannot, in general, be done independently, as the force autocovariance depends on both the time $\ti$ as well as the control parameter $\lambda_i$.

\section{\label{appendix:simulation_details}Simulation details}
For equilibrium simulations, we describe the system's time evolution with an overdamped Langevin equation
\begin{equation}
\frac{\md x}{\md t} = -\beta D \partial_{x}U(x,\lambda) + \sqrt{2D}\xi(t) \label{appendix:overdamped_langevin} \ ,
\end{equation}
for system position $x$, inverse temperature $\beta \equiv (k_{\rm B}T)^{-1}$, diffusion coefficient $D$, force $-\partial_{x}U(x,\lambda)$, and zero-mean white noise process $\xi(t)$ satisfying $\langle\xi(t)\xi(t')\rangle = \delta(t-t')$.
	
Given the potential energy~\eref{periodic_potential} with underlying period $\ell$, we calculate the conjugate force autocovariance $\langle\delta f(0)\delta f(\tgen)\rangle_{\lambda}$ over a discrete mesh of control parameter values $\lambda \in [0,\ell)$ and lag times $\tgen\in(0,\infty)$, by simulating system dynamics at fixed control parameter. We estimate the autocovariance between mesh points with a bivariate cubic spline interpolation on the empirical $(\lambda,\tgen)$ mesh.
	
After obtaining the force autocovariance from equilibrium simulations, we numerically minimize the excess work~\eref{cost_function} subject to the constraints of fixed duration $\tau$~\eref{protocol_duration} and endpoints $\lambda_0,\lambda_N$. In particular, we implement a Sequential Least SQuares Programming (SLSQP) algorithm (also known as Sequential Quadratic Programming)~\cite{winston} provided by the \texttt{scipy.optimize.minimize} python package using the \texttt{SLSQP} option. 

\end{document}